\documentclass[12pt,twocolumn]{aastex6}
\usepackage{natbib,amsmath,graphicx,epsfig,amssymb}
\usepackage{url}
\usepackage{multirow}
\usepackage{xspace}

\newcommand{\ba}{\begin{eqnarray}}
\newcommand{\ea}{\end{eqnarray}}
\newcommand{\be}{\begin{equation}}
\newcommand{\ee}{\end{equation}}
\newcommand{\codestyle}{\tt}
\newcommand{\Patch}{{\codestyle PATCHWORK}\xspace}
\newcommand{\HARM}{{\codestyle HARM3D}\xspace}

\begin{document}

\title{\Patch: A Multipatch Infrastructure for Multiphysics/Multiscale/Multiframe
Fluid Simulations}

\author{Hotaka Shiokawa\altaffilmark{1,2}, Roseanne M. Cheng\altaffilmark{2,3}, Scott C. Noble\altaffilmark{4,5}, Julian H. Krolik\altaffilmark{2}}

\affil{\altaffilmark{1}Harvard-Smithsonian Center for Astrophysics, 60 Garden Street, Cambridge, MA, 02138, USA}
\affil{\altaffilmark{2}Department of Physics and Astronomy, Johns Hopkins University, Baltimore, MD, 21218, USA}
\affil{\altaffilmark{3}Computational Physics and Methods Group, Los Alamos National Laboratory, P.O. Box 1663, Los Alamos, NM, 87545, USA}
\affil{\altaffilmark{4}Department of Physics and Engineering Physics, University of Tulsa, Tulsa, OK, 74104, USA}
\affil{\altaffilmark{5}NASA Postdoctoral Program Senior Fellow, Goddard Space Flight Center, Greenbelt, MD, 20771, USA}

\begin{abstract}
We present a ``multipatch" infrastructure for numerical simulation of fluid problems
in which sub-regions require different gridscales, different grid geometries, different
physical equations, or different reference frames.  Its key element is a sophisticated
client-router-server framework for efficiently linking processors supporting different
regions (``patches") that must exchange boundary data.  This infrastructure may be used
with a wide variety of fluid dynamics codes; the only requirement is that their primary
dependent variables be the same in all patches, e.g., fluid mass density, internal energy
density, and velocity.  Its structure can accommodate either Newtonian or relativistic
dynamics.  The overhead imposed by this system is both problem- and computer cluster
architecture-dependent.
Compared to a conventional simulation using the same number of cells and processors
employed on a problem not requiring multipatch methods, the cell-update per processor
rate decreases by an amount that can range from negligible to a factor of a few; however,
even in these problems, the infrastructure can permit substantial decreases in the
total number of cell-updates required.

\end{abstract}

\keywords{methods:numerical --- hydrodynamics --- MHD}

\section{Introduction} \label{sec:intro}

\subsection{Importance of multiphysics/multiscale/multiframe capability} \label{subsec:multiphysics/scale}

    Many important physical processes involve heterogeneous systems in which the nature of the matter in different regions exhibits strong contrasts. The material may vary in its characteristic internal length or time scales, or in its local geometric symmetry.  There may even be contrasts in which the physical mechanisms of importance differ between regions: for example, chemical reactions or self-gravity may be significant in some, but not all locations.  These regions may also move with respect to one another, perhaps changing shape as they do.  When the regions have relative motion, the fact that physics is often most concisely described in a system's mean rest-frame means that no single rest-frame is appropriate for the entire problem.  At the same time, interactions between these regions may nonetheless demand simulation methods allowing data from one region to inform the behavior of another.

   Problems exhibiting strong contrasts in length or time scales are called ``multiscale problems".  We will also use this term to include contrasts in grid symmetry.  In multiscale problems, numerical methods work best with different grid systems in different regions, perhaps contrasting in resolution, perhaps in symmetry, e.g., polar vs. Cartesian.   Those involving disparities in mechanisms are called ``multiphysics problems".  In these problems, one must solve entirely different equations: those of magnetohydrodynamics (MHD) rather than those of hydrodynamics, or with or without transport processes such as viscosity or diffusion.  Problems with internal frame shifts we dub ``multiframe problems".  For these, it would be desirable to translate the equations from one frame to another in different portions of the calculation.  Astrophysics is rich in problems to which at least one, and sometimes all, of these labels apply, and therefore at least one, and sometimes all, of the difficulties, both technical and conceptual, that they pose.

    To illustrate their significance, consider a few examples.  The topic that initially motivated our work is the mechanics of accretion around a binary system.  For us, the partners in the binary are supermassive black holes (see, e.g., \citealt{Schnittman13}), but much about this problem changes little whether the binary comprises a pair of proto-stars (e..g, as imaged and analyzed by \citealt{Mayama10}) or a pair of black holes.   In this situation, there are widely disparate scales because the structure of the circumbinary disk varies on the scale $a$ of the orbital separation, whereas most of the accretion power emerges at the inner edges of the disks orbiting around the individual masses (often called ``mini-disks"), which could be a great deal smaller.  In addition, throughout these disks, even to define the saturation level of the MHD turbulence supplying the accretion torques requires treatment of lengthscales small compared to the disk scale heights, which could be considerably smaller than the radial scale.  There are also differences in the symmetry of well-designed local grids.  Because angular momentum transport in accretion disks is slow compared to the orbital time, it is very important that there be little numerical momentum diffusion; this fact demands a grid mimicking the symmetry of their nearly-circular flow (see, e.g., \cite{Sorathia2013}).   However, such a grid would be polar and centered on the binary center-of-mass for the circumbinary disk, whereas for each mini-disk, it would be polar and centered on the object whose gravity is most important for that disk.  A single Cartesian grid for the entire system would likely produce an intolerable level of numerical diffusion.   This binary accretion problem is also one that demands multiple reference frames for much the same reason it requires multiple sub-grids with different symmetries.  The physics of the circumbinary flow is easiest to grasp in the center-of-mass frame; that of the individual mini-disks in the frame of each member of the binary.  Thus, one would like to be able to divide this calculation into at least three different zones, each with its own grid and reference frame.

Another example may be found in tidal disruption of stars by supermassive black holes, which has become a subject of great interest in recent years as numerous examples have been found (in both optical/UV, e.g. \citealt{Gezari2009} and \citealt{Arcavi2014}, and in X-rays: \citealt{Auchettl2016}).  This is a multiscale problem because it is necessary both to resolve dynamics within the star as it is broken apart and to follow the fluid dynamics of the debris as it gradually accretes onto the black hole.  Measured in terms of gravitational radii $r_g$ defined relative to the mass $M$ of the black hole ($r_g \equiv GM/c^2$), main sequence stars are $\sim 1 M_6^{-1} r_g$ in diameter, where $M_6$ is the black hole in units of $10^6 M_\odot$.  Thus, to follow their break-up requires cells $\ll 0.1 M_6^{-1} r_g$ in size.  On the other hand, the debris orbits have semi-major axes $\sim 10^3 M_6^{-1/3} r_g$, so that the fluid motion after stellar break-up takes place on a much larger scale.   Nonetheless, despite this dramatic scale contrast, the break-up of the star is inextricably tied to the much larger-scale debris motion. It is also a multiphysics problem because stellar self-gravity, not surprisingly, is of the essence so long as the star stays in one piece, but after its matter is spread sufficiently widely, it becomes inconsequential.   And it is a multiframe problem because the mechanics of a nearly hydrostatic star are definitely best viewed in the star's frame where the fluid velocities are small, whereas the mechanics of an accretion flow are far more easily understood in the black hole frame.  Its multiframe nature also creates a contrast in grid symmetry because coherent stars are most naturally treated in a spherical coordinate system whose origin is the center of the star, whereas flow around a black hole is best described in a cylindrical or spherical coordinate system whose origin is the center of the black hole. Thus, this problem, too, involves all these categories of complication.   In fact, we have chosen it as the subject of our first test-problem for our new infrastructure.  We will quote a few technical results from this test-problem here; a full analysis will be published separately.

\subsection{State of the art and its limitations}

The desirability of overcoming these challenges has not gone entirely unnoticed by the computational community, and a number of partial solutions have been developed. Adaptive Mesh Refinement (AMR) methods can dynamically adjust spatial resolution to follow local lengthscales \citep{BergerOliger1984,BergerColella1989}; a closely-related scheme, overlapping moving grids, can be used to follow a coherent region with a distinct spatial scale or symmetry.  MultiProgram/MultiData (MPMD) methods \citep{Barney} offer a convenient way to evolve different regions according to the different mechanisms acting in them\footnote{In practical terms, users run multiple---possibly different---executables each producing a different data product and all sharing the same Message Passing Interface (MPI) ``MPI\_COMM\_WORLD" communicator.   Typical, i.e. single-program, parallelized calculations are launched using the same executable on all processors.  Since some codes (e.g., \HARM) set algorithmic choices (e.g., coordinate system type) at compile time, MPMD allows users of such codes to run different algorithms on different sets of processors and still allow all of the processors to communicate with each other.}.  We will call these separate regions ``patches", hence the name ``multipatch" for our general approach.  Because each patch is run by an independent program under MPMD, their only interaction is through the exchange of boundary conditions.

For some of these methods, professionally-supported implementations are available.  {\codestyle Chombo}, for example, is a particularly well-developed AMR package \citep{Chombo}.   It permits the use of two different methods to divide up regions into separate grids, embedded boundaries and mapped multiblocks.  There are also relativistic versions of these fixed multiblock methods \citep{Clough2015,Schnetter2014}.  General relativistic hydrodynamics has been treated \citep{Blakely2015} using the techniques of {\codestyle Overture} \citep{Overture}, which offers users the option of moving overlapping grids.  In this approach, a composite grid is formed in order to bridge two contrasting overlapping grids, with data from each interpolated onto the composite grid \citep{ChesshireHenshaw1990}.  Numerical relativity calculations can use multiblock infrastructure with AMR, but all dependent variables must be defined in terms of a global Cartesian tensor basis \citep{Llama}. Similarly, there are numerous MPMD systems permitting computation of different physics in different parts of a global system.   These range in their applications from linking multiscale fluid simulation to molecular dynamics \citep{Nie2006} to modeling blood flow through the brain \citep{Grinberg2013}.

Another approach to solving the problems of multiple scales, but not multiple physics, is the use of moving unstructured grids (e.g., the codes {\codestyle AREPO}: \citealt{Springel2010} and {\codestyle TESS}: \citealt{Duffell2011}).  Schemes like these very flexibly place resolution where it is required for the hydrodynamics. It has also recently become possible to extend them from hydrodynamics to magnetohydrodynamics \citep{Duffell2016,Mocz2016}. They do not, however, naturally retain the virtues of conforming to natural symmetries of the problem (e.g., suppressing numerical diffusion by aligning cell axes with the fluid velocity), nor do they readily permit the use of contrasting physics in different regions.  With significant effort, it is possible to avoid the first drawback \citep{Duffell2016}, but a new solution must be created for each new problem.

Despite the real successes of all these different schemes, there remain significant barriers to their employment on many kinds of problems.   Multiblocks must fit smoothly against one another in a fixed configuration, while embedded boundaries require Cartesian grids.  Neither of these allows relative motion of the cell blocks.  Most importantly, none of the methods introduced so far achieves the simplification and efficiency gains that arise from following moving regions' physics in their own reference frames.

The advantages of working in the most suitable reference frame can be substantial.  Consider, for example, a hydrodynamics problem in which structure A, containing only motions subsonic relative to its own center-of-mass and varying on short lengthscales, moves supersonically within a larger background fluid B with longer gradient scales.    If such a problem were treated with a moving grid scheme, the time-step within region A would be severely limited by its supersonic velocity and its small cell sizes; transformation to the moving frame could reduce the number of time-steps required by a large factor.  Numerical accuracy would also be substantially improved as there would be no need to perform numerous close subtractions of velocities in order to find the relative velocities between cells.

Analogous advantages can stem from equation simplification.   Suppose, for example, that in the moving region there is a diffusive transport process that is unimportant in the background.    Treating this transport process in the moving frame eliminates what would otherwise be a large, unnecessary, advective flux.   If the velocity contrast between the regions approaches the relativistic level, treating everything in the background frame introduces serious conceptual problems: because classical diffusion causes instantaneous transmission of information \citep{MF1953,Narayan1992}, such problems cannot be formulated covariantly.  Transformation to a frame in which local motions are slow permits a clean use of the local Newtonian limit in which diffusive transport is mathematically consistent.

The limitations of existing methods severely crimp study of many interesting problems, including the two we mentioned as motivating our work, the dynamics of accretion flow in binary systems and tidal disruptions.  In the former case, none of the existing systems supports the optimal grid geometries, a pair of small spherical grids moving with respect to a larger (and coarser) spherical grid whose origin (the system center-of-mass) is stationary.  Instead, one would be forced to cover a large region with small cells, none of them aligned with fluid motions and therefore incurring large numerical diffusion.  Likewise in none of them does one gain the advantage in numerical accuracy afforded by working in a locally co-moving reference frame.  It is also difficult in this framework to avoid much of the simulation being burdened by a very small time-step relevant in only a sub-region.  Simulations have been attempted, but they have been either restricted to 2D \citep{RyanMacFadyen2017,Bowen2017} or, if 3D, limited to very short duration \citep{Bowen2018}.
The restrictions are even more severe for the tidal disruption problem. A number of simulations have been carried out whose grid origins follow the center-of-mass of the star
\citep{Cheng2013,Cheng2014,Guillochon2013,Guillochon2014}, but they cannot continue to follow the career of the tidal debris during the much longer time that it orbits the black hole because their high-resolution grids are extremely inefficient for that problem.  It is also important to compute the star's self-gravity while it survives and before its tidal debris disperses, but then turn it off afterward for efficiency, yet no existing system provides the flexibility to ignore the regions where self-gravity is unimportant. These two astrophysical problems demonstrate the need for greater flexibility, but it is easy to imagine many other problems for which current methods are inadequate.

\subsection{Our innovation}

The system we present here, which we call \Patch, is designed to eliminate the limitations due to use of a single reference frame for all patches, while also maximizing ability to simultaneously deal with issues of multiple scales, multiple grid symmetries, and multiple varieties of local physics.  Exploiting the flexibility of the MPMD approach, it utilizes well-defined coordinate transformations informed by relativistic methods (but not restricted to relativistic problems) to simulate heterogeneous systems in which regions requiring independent treatment are regarded as independent processes operating in independent reference frames.  Each patch has its own grid, with its own resolution scale and symmetry; among the many benefits offered by independent local coordinate systems, patches give an easy solution to the problems raised by coordinate singularities (e.g., at the origin of a polar system). Each patch also solves its own equations, whatever is necessary to do the job in that region.

The relationships between these frames are defined by coordinate transformations with the ability to eliminate local mean velocities, so as to reap the benefits just described, while retaining a common overall time so that the entire simulation can advance together.  This approach creates an important simplification---the same coordinate transformation that relates array index-space to spatial coordinates can also be used to eliminate bulk velocities.  This single coordinate transformation also applies to the metric with respect to the coordinates. In addition, provided that the physical quantities entering into the boundary conditions are scalars, vectors or rank-2 tensors, their transformation from the coordinate system of one region to that of another is well-defined and straightforward.  In this way, the free use of arbitrarily curvilinear and arbitrarily discretized coordinates can be combined with the virtues of treating local physics in its most natural reference frame.

However, in order for the patches to exchange boundary conditions at simultaneous times, the time coordinate in all the patches must be the same.  In relativistic terms, this means that the coordinate transformations relating patches with relative motion are {\it not} Lorentz transformations; as a result, the reference frames of moving patches are in general non-inertial.  This policy may be somewhat unfamiliar, but because the equations of physics can all be written in completely covariant fashion, their form under this sort of transformation is well-defined.

Our version of the multipatch system also offers several additional features.  Because the patches interact only through boundary condition exchange, they can have independent time-steps; because long time-step patches need many fewer updates to traverse the same physical time, when parallelized those patches can be assigned many more cells per processor to achieve better load-balancing.  In addition, patches can be added or removed from time to time as conditions change and different demands arise.

To mitigate the complexity and overhead created by inter-patch communications when the computation is parallelized, we have created a client-router-server system that efficiently links the correct processors in each patch to their boundary condition partners in other patches.

Lastly, our package is structured as a ``wrapper" to fit around a
user-supplied hydrodynamic or magnetohydrodynamic simulation code.
The only requirements placed upon these codes are that they
  should have fixed grids, and their primary dependent variables (the
ones whose boundary conditions are exchanged between patches) are all
the same.   In intrinsically conservative codes (e.g.,
    \HARM, our test code), boundary conditions are generally fixed
    through the ``primitive variables'' (density, velocity, internal
    energy) because the Riemann problem is defined in terms of them.
    Other algorithms may lead to other choices and \Patch should
    accommodate them.  Here, we illustrate its performance only
    through \HARM.

\subsection{Outline of the paper}

In Sec.~\ref{sec:method}, we set out the principal features of our method.  This presentation begins with an overview (Sec.~\ref{subsec:method_code_structure}) in which we define what we mean by a ``patch" and describe how different patches are related to one another.  The next subsection discusses the principal operation of the multipatch system, boundary condition exchange between neighboring patches.  Following that, in Sec.~\ref{subsec:method_interpolation}, we detail how boundary data are interpolated from one patch's grid to another's.  Sec.~\ref{subsec:method_add_remove} briefly discusses how users can add new patches and remove old ones.  Then Sec.~\ref{subsec:method_parallel} presents the method's core: the client-router-server architecture we created so that, in a parallelized environment, a processor in one patch is linked to the correct partner in a different patch in order to exchange boundary data.  The final subsection, Sec.~\ref{subsec:method_timestep}, explains how different patches can advance with different time-steps and yet remain synchronized.  We lay out in the Appendix an overview of how these operations are organized into discrete routines, and fluid codes can, with a small number of additional lines of code, be made compatible with the \Patch system.

Sec.~\ref{sec:tests} presents a variety of tests of our multipatch implementation.  In its first subsection, we demonstrate that a shock can pass smoothly from one patch to another without alteration.  In the second, we show that even when patch symmetries contrast strongly, a blast wave can travel from one to the other and remain close to the Sedov-Taylor similarity solution.  In the third, we examine the degree to which interpolation of data from one patch to another may lead to departures from rigorous conservation of mass, momentum, and energy, and discuss how such departures can be kept small.

Sec.~\ref{sec:efficiency} discusses computational efficiency by presenting benchmarking tests and scaling data for the overhead imposed by the multipatch system. 

The last section summarizes the paper.

\section{Method}
\label{sec:method}

 We have designed our multipatch software to be compatible with any numerical simulation code
in which the computational domain is discretized into a fixed set of discrete grid-cells, and the
primary dependent variables, the ones exchanged between patches as boundary conditions, are the
same in every patch. It also requires the geometric factors relevant to the operators of vector calculus to be defined in terms of metric elements; this is done as a matter of course in relativistic codes, but it is also a feature of a number of contemporary Newtonian codes such as Athena++ and the most recent version of Zeus \citep{Athena++,Sorathia2013}. For development and testing purposes, we have used it with the finite volume general relativistic hydrodynamics code \HARM \citep{Noble2009} running in every patch.   In the near future, we plan to port it to other codes to demonstrate its flexibility.  We expect that, subject to the stipulation about variable consistency, it will be possible for different codes to run in different patches.

\subsection{Overview}
\label{subsec:method_code_structure}

\Patch's structure is based on the concept of ``patches".  A patch is
a region of space defined by the user.  Locations within it are
described by its particular coordinate system and discretized according
to its own particular grid.   The time-evolution of its fluid's
physical properties is governed by a particular set of equations,
always including the Euler fluid equations, but potentially
extensible to the Navier-Stokes equations or the MHD equations,
and potentially supplementable by chemical or nuclear reaction
networks, a Poisson solver for self-gravity, or other sorts of
equations.  Evolving an individual patch is the responsibility
of an individual process within the MPMD environment. Although it
is possible to run the simulation as a single program, using one
program for each patch keeps the code simple and conceptually clear.
As a result, the method is intrinsically parallelized: there must
be at least one processor for each patch.

\Patch coordinates a number of different individual patch processes
through the incorporation of several specific routines into the
fluid simulation code chosen by the user. Some are problem-independent,
but others are problem-specific and therefore need to be written by
the user.  The most important functions of the \Patch routines are to:
define the trajectory of each patch in terms of the ``background
coordinates" (see below for definition); calculate the coordinate
transformation matrices necessary to translate physical quantities
and locations between each patch and the background
coordinates; control boundary condition exchange between different
patches while maintaining ``situational awareness" about which portions
of a patch's boundary adjoin other patches and which are on the edge
of the physical problem volume; and synchronize the time-steps in the different patches.
In addition, there are several other optional multipatch-specific
routines that will be described later.

A system of ``background coordinates" underlies the entire region being simulated.
The locations and boundaries of all the patches are defined in terms
of this system.  It is always Cartesian, and its time coordinate is the
universal time for all the patches' coordinate systems. Its purpose is
both to serve as a reference for positions and to serve as a ``common
language" for all patches to describe the locations of exchanged data.

Individual patches can have any shape or size,  provided only that they
fit within the background coordinate grid.  They can be stationary relative
to the background coordinates or move.  Their internal coordinate systems and
grids are entirely independent of all other patches' spatial coordinate systems
and grids.  It is convenient in many problems to divide the patches into two
categories, ``global" and ``local".  Frequently, one patch provides
the great majority of boundary condition data for the other patches
and occupies all or a large part of the problem volume.  When that
is the case, that patch is deemed ``global", and its reference frame
is tied to the frame of the background coordinates. Its internal spatial
coordinate system, however, can still be defined independently of the
background coordinates.  Although it is often convenient to have a
global patch, it is not a requirement of the system.  Any patch not designated
as ``global" is considered to be ``local".

When two or more patches overlap
in their spatial coverage, only one of them governs the dynamics within
the overlap volume. We then speak of the ``active patch" updating the
properties of the ``uncovered cells" and the ``inactive patch" containing
the ``covered cells".  If one patch is a local patch and the other is
a global patch, the local patch is always the active one.  If two or more
local patches overlap, the user designates the hierarchy of activity in advance.
As covered cells approach the patch boundary of the active patch, they
become ghost cells for other cells in their patch that are already
uncovered.  At that point they are filled by interpolation from the active
patch cells covering them.

One spacetime is specified for the entire problem volume with a metric
defined on the background coordinates.  This spacetime can be described in any
of the patches by means of the appropriate coordinate transformation from the
background system to the patch coordinate system.
In some instances, the fluid mass in one or more of the patches may be important to gravity
throughout the problem volume.  The best way to account for this contribution to gravity
depends on circumstances.  Relativistic velocities or strong gravity demand solution
of the Einstein Field Equations.  Because these equations, like the hydrodynamics
equations, are hyperbolic, they can also be solved within the multipatch framework.
On the other hand, if the fluid moves more slowly and the gravitational field is
weak, the Poisson Equation is appropriate, which is elliptic.  This case requires the
global patch physics repertory to include Newtonian self-gravity, and the density distribution
from any local patch with significant mass must be interpolated to the global grid,
although possibly with crude resolution.

Physical consistency likewise demands that all patches are updated
according to the same time coordinate and must reach a given value
of this time coordinate together.  Such synchronization is achieved
automatically if all advance with the same time-step.  However, as
we discuss below (Sec.~\ref{subsec:method_timestep}), this is not necessary.
If some patches can be evolved stably and accurately with a longer time-step
than others, it is necessary only for the patches all to be synchronized after
one time-step of the patch with the longest step.  Note that because
there is a single time coordinate for all patches, the coordinate
transformations between them are {\it not} Lorentz transformations
unless the relative velocity between the two patches being linked
is zero and both patches are inertial.

In the course of each update, patches bordering on one another must
exchange boundary condition data.  Accomplishing this step is the
core of our system. 

\subsection{Boundary condition exchange between patches}
\label{subsec:method_bc}

Within any particular patch, we distinguish three types of boundary zones
for individual processors.  Ghost zones covered by other processors in the
patch are in the first category.  The second category comprises ghost zones
lying on the physical boundary of the problem.  The third category is of
greatest interest to the multipatch scheme, those ghost zones covered by
processors assigned to other patches.

\begin{figure}[ht!]
\figurenum{1}
\plotone{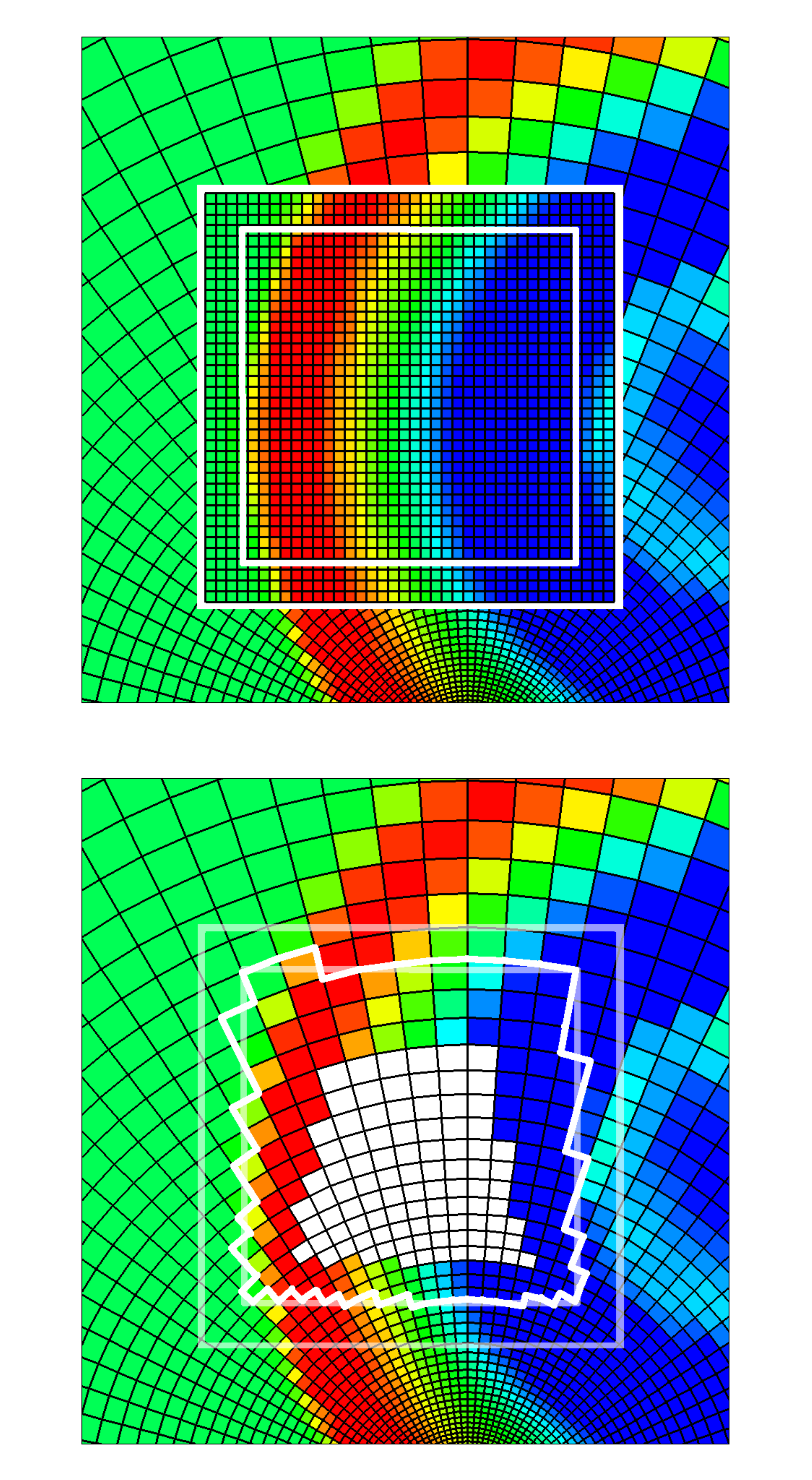}
\caption{
A snapshot of internal energy density (color contours) and grid-cells
in a 3D blast wave simulation.
(Upper panel):  White squares show the physical boundary (inner) and numerical boundary
(outer) of the local patch.  Where the local and global patches overlap, only the
local grid is shown.  (Lower panel): Like the upper panel, but where the patches
overlap, only the global grid is shown. The colored cells inside the jagged loop
are filled with data interpolated from the local patch to the global patch.
The white cells inside the jagged loop are unused when the patches are in this
configuration because the local patch updates the physics in their volume.
The physical (inner) and numerical (outer) boundaries of the local patch
are shown as thin white lines for reference.
\label{fig:ghost}
}
\end{figure}

The first two can be handled by the standard devices found in
existing fluid codes.  Here we describe
how the boundary information is obtained for ghost zones in the third
category.  We begin by displaying an example so that readers can easily
visualize the issues involved (Fig.~\ref{fig:ghost}).  In this figure,
the fluid's internal energy density is represented by color contours and grid cells are
delineated by black lines.  We have chosen to follow the fluid mechanics
in this example by means of two patches, a finely-resolved Cartesian
local patch and a more coarsely-resolved polar global patch whose radial
grid is logarithmically-spaced.
In the upper panel of the figure, the physical boundary of the
Cartesian patch is shown by the inner white box; the area covered
by its ``ghost zones", the cells needed to establish boundary
conditions for the physical region, lies between the two white boxes.
The lower panel shows the converse situation: the jagged white
contour shows the boundary of territory in the global patch not
covered by physical cells of the local patch; the cells between
that jagged contour and the white cells are where the global patch
needs boundary data.

\begin{figure}[ht!]
\figurenum{2}
\plotone{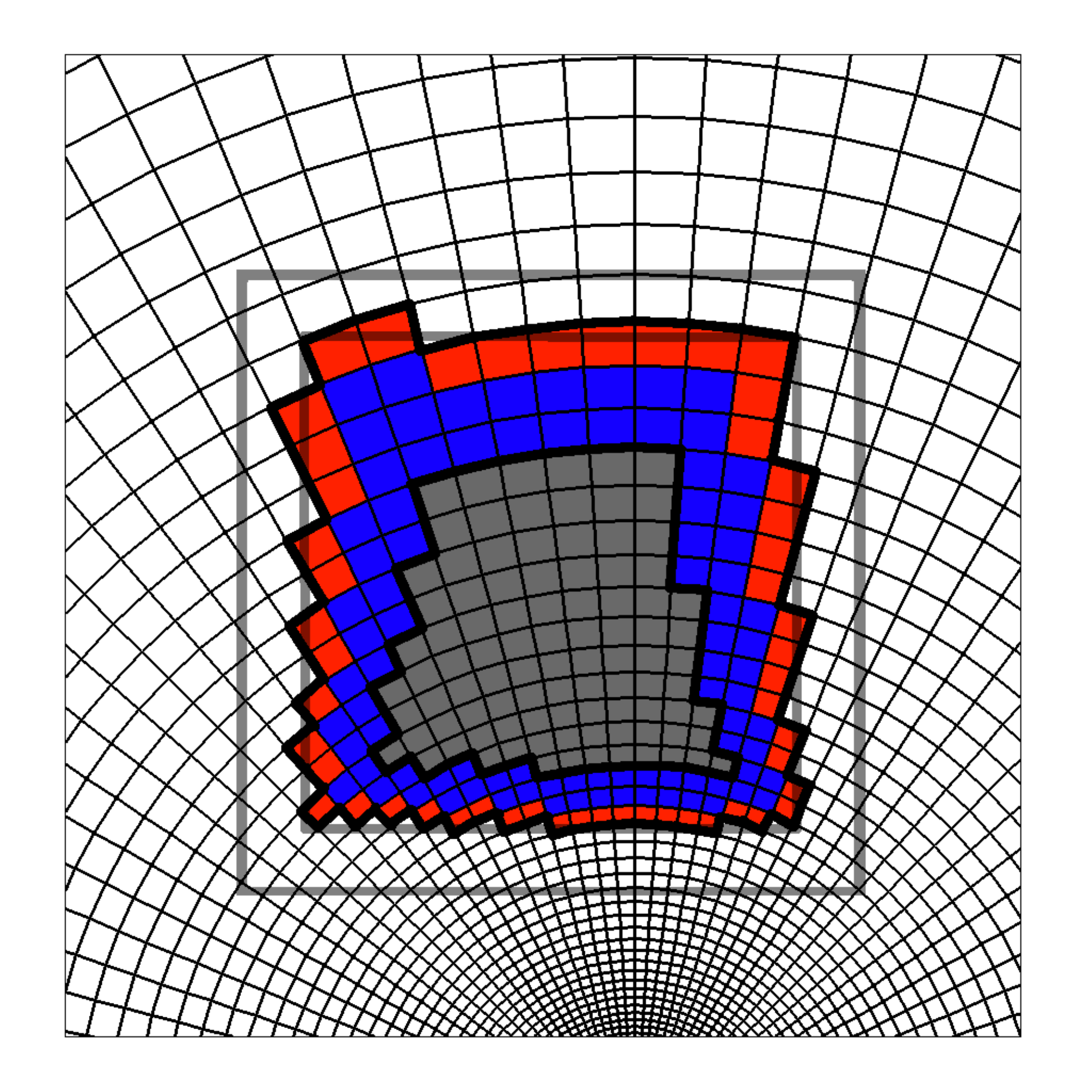}
\caption{
The ``flags" assigned to the global grid-cells for the same snapshot shown in
Figure~\ref{fig:ghost}.  Red and blue indicate inner and outer ghost-zones (for
the global patch), respectively.  White cells are ordinary cells in the
global patch interior; gray labels global cells covered by the local patch
and ignored.  The thin gray squares show the physical (inner) and numerical (outer)
boundaries of the local patch.
\label{fig:flag}
}
\end{figure}

The first step is to discover which processors in which patches have
the information.  To minimize inter-patch communication time, we organize
this process to avoid exchanging unused data.  Because this procedure is
almost independent of whether the patch needing boundary data is a global
or a local patch, for this part of the discussion we call them ``patch~A"
and ``patch~B".  We begin by labeling all the zones in patch~A (here
this happens to be the global patch) with an integer array, illustrated in
Figure~\ref{fig:flag}.  The values in this ``flag array" denote whether a
zone is a ghost zone, and if so, what type of ghost zone.  This array must
be updated at each time-step if any of the relevant patches move (at each
synchronization time-step in the case of heterogeneous time-steps:
Sec.~\ref{subsec:method_timestep}).  In the
figure, the white zones are in the interior of patch~A, and have nothing
to do with boundary conditions.  Gray zones are the zones in patch~A completely
covered by patch~B; they, too, are irrelevant to boundary
conditions\footnote{If patch~A were a local patch, it would have gray zones
only if patch~B were another local patch, and patch~B took priority over
patch~A; we have not yet implemented ``local-local" boundary data exchange,
but plan to do so soon.}. A zone in
the global patch is considered to be covered by the local patch if its
center falls within the local patch's physical region. The red and blue zones
are covered cells in patch~A that act as ghost-cells for uncovered cells
in patch~A.  The red cells directly touch uncovered patch~A cells.  Fluxes across
their inner (in a topological sense) faces are used in updates of  the uncovered
cells they touch.  Blue zones are the outer layer of ghost-cells needed for updates
of uncovered cells in patch~A; in \HARM the ghost-cell zone is three cells wide,
so the blue cells are either the second or third ghost-cell from the last uncovered
patch~A cell along at least one dimension.  They are used in the internal
reconstruction by which cell-center values of fluid quantities are extrapolated
to the face touching the physical boundary.
As long as the number of ghost
cells is adjusted appropriately, any reconstruction method should in
principle work with \Patch.  In the tests presented here, we
used piecewise parabolic reconstruction \cite{1984JCoPh..54..174C} 
with a MC (monotonized central-differenced) slope limiter.

It is important to note that this system is thoroughly agnostic about many
of the possible choices made in different codes.  Because the coordinates at
which the boundary data are needed are determined by the fluid code operating
in the requesting patch, it doesn't matter whether that code defines the
variables at cell-centers or the centers of cell-faces or anywhere else; it
knows the locations at which it needs the information, and it is the job
of the responding code (which may be an entirely different one) simply to
interpolate its data, no matter how defined in terms of location within cells,
to the proper point.   The system is even
capable of accommodating codes with different numbers of ghost-cells.
\HARM, for example, requires three layers of ghost-cells, but \Patch contains
a parameter that can be set to whatever number of layers the user's code needs.

Once patch~A determines the locations of its ghost zones' centers, a list of
the background coordinates for these locations is sent to all other potential patch~B's
along with a request for interpolated values of the fluid variables at the coordinate locations.
Patch~B interpolates within its grid in order to find the values at
the locations desired by patch~A.  It then transforms the data from its
coordinate system to the background coordinate system, using the coordinate
transformation Jacobian linking patch~B to the background system.  Only then are
the boundary data transmitted back to patch~A, which transforms it from the
background system to its own coordinates. This procedure enables every
patch to deal with the incoming coordinate list independently, without knowing
anything about other patches.  Doing things this way is especially important
when patch~B moves.

Note that if patch~A is the global patch, the first step is done differently.
At initialization, the local patches are informed of the cell locations at which
the global patch dependent variable data are defined.  Because the local patches
know their own positions in terms of background coordinates, they can determine
on their own what data the global patch needs.  This alternate procedure has
the virtue of diminishing inter-patch data transmission.

\subsection{Interpolation}
\label{subsec:method_interpolation}

Although remarked on only briefly in the preceding sub-section, there are
a number of subtleties to data interpolation, and multiple mechanisms may
be used. In the current version of our system, we use a comparatively
simple method, but this could readily be upgraded to something more
sophisticated for problems requiring it.

In principle, an arbitrary number of zones could be used to support
interpolation to a single point.  However, it is generally best for
the interpolation stencil to extend away from the point by a number
of zones that is no more than the number of ghost-zone layers
(usually 2 to 3), so that the stencil does not extend into another processor's
domain.

For our current method, we employ tri-linear interpolation.  We locate
the grid corner closest to the interpolation point and define the stencil in 3D
to be the centers of all eight cells touching that corner.  This method
works quite well when the dimensions of the cells in patch~A and patch~B
are comparable (see Sec.~\ref{subsec:conservation}), but can lead to
errors when they are not.  In some sense, this is unsurprising: if
there is structure on the finest scale supported by one of the patches,
it cannot be well-represented by a much coarser grid in the other.
However, the trouble can also move in the opposite direction because
the eight cells in the finer grid nearest the interpolation point may
together cover only a small part of the volume of the ghost-zone in
question if its grid is much coarser.  Sometimes errors of this latter
variety can be substantially reduced by replacing the values in the inner layer of ghost-cells
with a wider average over nearby cells.
Such an operation effectively magnifies the volume of the finer-scale
grid contributing to the coarser-grid ghost-cells.

Without special methods, interpolation does not necessarily conserve quantities.
To achieve strict mass (or momentum or energy) conservation in our data interpolation
could require identifying all cells that fall within the ghost-cell
and summing their contributions.  If the ghost-cell boundaries cut
obliquely (or even worse, in a curve) across some of the interpolation
cells, one would need to adjust their volumes accordingly.  Although this is
possible if both global and local patches are in Cartesian
coordinates, it becomes a non-trivial mathematical problem once any of
the patches are in curvilinear coordinates.  In Sec.~3.3 we test quantitatively
how closely our interpolation method comes to conserving mass and momentum.

\subsection{Adding and removing patches}
\label{subsec:method_add_remove}

Stationary or moving patches can be added or removed throughout
the simulation anywhere within the physical problem volume.  This is done
using the flag array for the ghost-zones discussed in Section~\ref{subsec:method_bc}.
Although these flags are most often used to signal the need for data interpolation
from overlapping patches, they can also be used to signal the need to interpolate
data for other reasons as well---such as removing or introducing a new patch.
To remove a local patch, one temporarily changes the flags on all the zones
in the global patch covered by the local patch to ``ghost
zones" so that all of them are filled with interpolated data provided
by the local patch.  Once these zones are filled with data, one changes the
flags back to their normal state.  To add a local patch, one creates a
new patch process and to define its initial condition sets all its cell flags
to indicate they are ghost zones.  Just as in the patch removal operation,
these cells are then filled with the data they need, and the flags can
be reset to normal as soon as that is done.  However, the simulation must
be stopped immediately after a patch removal or immediately prior to a patch
addition because either one demands a new domain decomposition for
processor assignment.

In principle, patches could be added or removed while running.
To do so, however, requires having a clear criterion for when to make
the change, a specific plan for the reallocation of processors because MPMD
does not permit any change in the total number of processors while running, and
synchronization of the resumption of fluid updates between all
processors and patches.  For the time being, we have not implemented
such a scheme.

\subsection{Parallelization and inter-patch communication}
\label{subsec:method_parallel}

One of the most difficult tasks in developing a multipatch code is
its parallelization.  It requires a sophisticated infrastructure
combining two levels of data communication.  In one, boundary data
exchange within a patch, a single executable exchanges information
between its multiple processors exactly in the way made familiar by
non-multipatch parallelized methods.  In the other level, boundary data
exchange between patches, it is necessary to enable effective data
communication when the pairing of processors with overlapping boundaries
evolves dynamically, and two independent executables,
both running within the MPMD environment, must be coordinated.

To describe how we achieve this, we first define a notation.  We label each
CPU in a simulation by ${C^i}_j$, where $i$ is a patch-ID and $j$ denotes
local CPU rank within that patch.  We set the patch-ID of the global patch to
$i=0$ and that of local patches to $i=1,2,...,N-1$, where $N$ is the total
number of patches in the simulation, including the global patch.  The index
$j$ runs from 0 to $n_i-1$, where $n_i$ is the number of
CPUs used for patch~$i$.

Consider a CPU at the edge of a patch, designated ${C^i}_0$.  This
CPU possesses boundary zones that need to be filled with
interpolated values.  It needs to know which CPUs ${C^k}_j$
in other patches handle cells lying under these zones (there could be
multiple values of $j$ satisfying this criterion, and sometimes
multiple values of $k$).  It
must also contact them to request interpolation values. The partner
CPUs ${C^k}_j$, on the other hand, need to know in advance that
other CPUs may be contacting them.  Because these relationships
constantly change if the patches move relative to one another,
this information must somehow be updated dynamically, even though
the patches may have differing time-steps.

To solve this problem, we construct a client-router-server system, setting up
inter-patch communication relationships that can persist throughout
the simulation.  In its simplest form, one CPU in each patch is chosen
to serve as the router, its liaison with all the other patches.  Then, when
client CPU ${C^i}_j$ needs information from beyond the boundary of patch $i$,
it transforms the coordinates of the cell-centers in question to
background coordinates and broadcasts that list to processors ${C^k}_{r_k}$,
where the $r_k$ processor in patch~$k$ is the designated router for
that patch.  The router processor in the $k$th patch then transforms the list from background
coordinates to patch~$k$ coordinates.  If all the cell-centers on the list lie outside
patch~$k$, the router replies accordingly.  On the other hand, if some
of them are inside patch~$k$, the router processor determines which of
the other processors working on patch~$k$ have responsibility for those
cells and distributes the request to those processors.  These processors, the servers,
interpolate their data to the correct positions, transform the results to background
coordinates, and return the results to the router. Finally, the router
transmits the information back to the client, CPU ${C^i}_j$.

\begin{figure}[ht!]
\figurenum{3}
\plotone{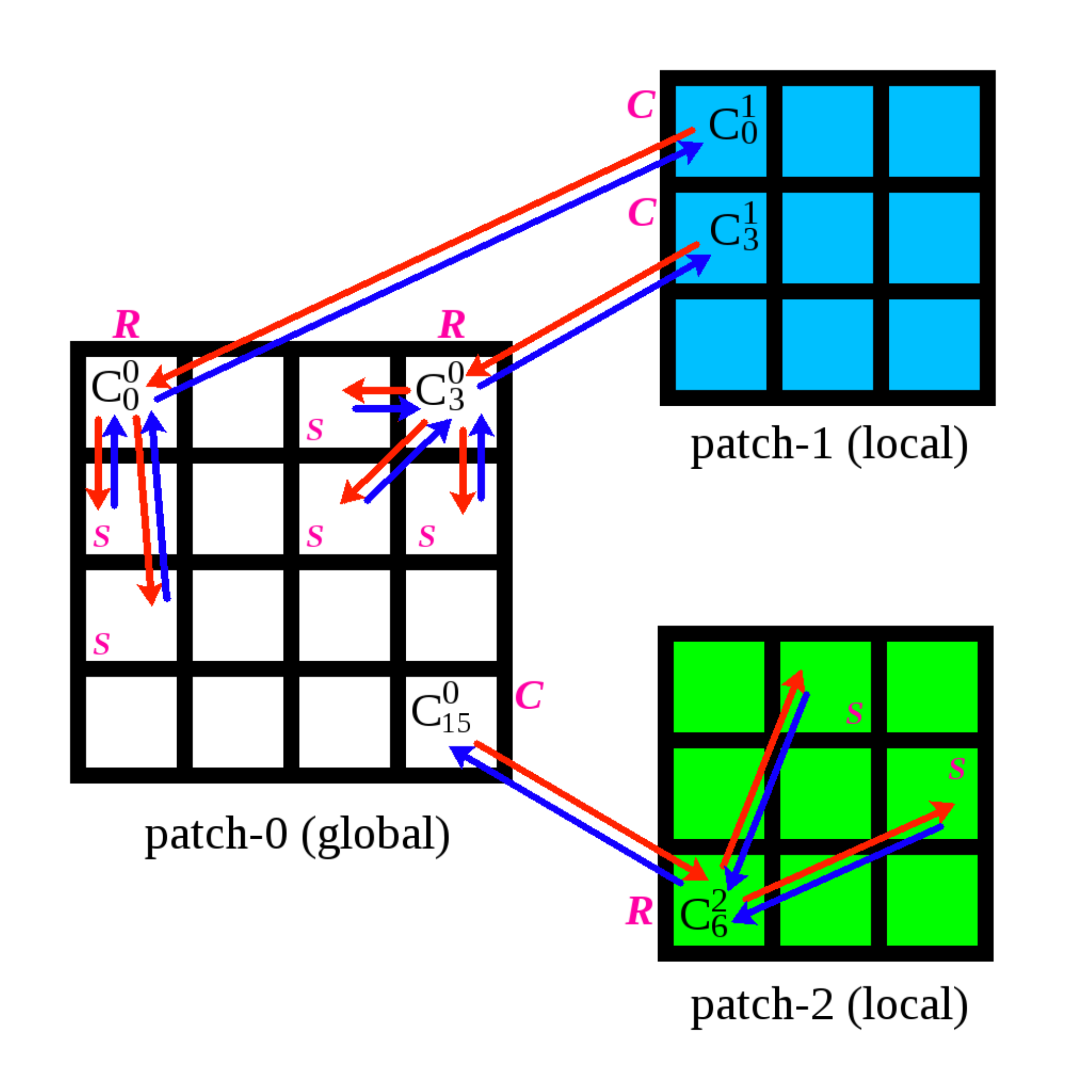}
\caption{
Schematic view of client-router-server relations for the multiple-router
scheme.  White, blue, and green patches represent the global patch (patch~0),
local patch~1, and local patch~2, respectively. Example clients, routers,
and servers are marked with C's, R's, and S's, respectively. Data requests
are shown by red arrows, data returned by blue arrows.  The local patches
reside inside the global patch, but they are placed outside the global
patch and enlarged for visualization of the information exchange system.
The squares in the patches represent CPU domains, not grid-cells.  In a
possible instance of data exchange, a client CPU in a local patch, ${C^1}_0$,
(upper left in patch~1) sends its list of ghost zones to its designated router
in the global patch, ${C^0}_0$, (upper left in patch~0).
${C^0}_0$ then communicates with appropriate server CPUs on its patch
to collect the requested data and returns the data to its client.  Simultaneously,
CPU ${C^1}_3$ also requests data from the global patch, working with its
global patch router ${C^0}_3$, which, in turn collects the information from
the relevant server CPUs and transmits it back to the client.  Even while
these two patch~1 CPUs communicate with their
partners in the global patch, it is possible for a CPU in the global patch, for
example ${C^0}_{15}$ (lower right in patch~0), to be a client, requesting data from
other patches such as patch~2; in this case, the router is ${C^2}_6$.
\label{fig:liaison}
}
\end{figure}

This communication scheme is conceptually simple
and easy to code. However, if only a single CPU is given router duties
for an entire patch, the communication load is shared very unevenly and
the great majority of processors sit idle while waiting for the routers to
finish their work.  To divide the workload more evenly, we regard {\it all}
processors as potential routers for their patch and redefine the
client-router relationship uniquely for each individual CPU (see Fig.~\ref{fig:liaison}).
These relationships are defined at the beginning of the simulation and
remain unchanged unless patches are added or removed.
For example, one may decide that ${C^1}_0$ always contacts ${C^0}_0$ for any information
regarding patch~0, ${C^1}_1$ always contacts ${C^0}_1$, and so on.
The function of the router is unchanged; it still determines which, if
any, of the processors on its patch holds the information requested and
acts as the go-between connecting clients and servers.

Although the varying numbers of processors per patch make an exactly even
division of labor impossible, a simple assignment scheme can spread it in
a reasonably even-handed manner.  If $C^{p_1}_i$ requires
information regarding patch~$p_2$, it contacts
\begin{equation*}
\text{router-}p_2( {C^{p_1}}_i ) =  {C^{p_2}}_{i \hskip -0.4pc \mod n_{p_2}}.
\end{equation*}
Note that CPUs on patch~$p_2$ could have 0 or multiple clients
on patch~$p_1$, depending on their index, $n_{p_1}$, and $n_{p_2}$.

\subsection{Heterogeneous time-steps}
\label{subsec:method_timestep}

One of the common problems in simulating multiscale systems with
grid-based hydrodynamics codes is that the time-step of the entire
computational domain is limited to a small value by a few regions
with small grid-cells and high characteristic fluid velocity.  As a result,
the remainder of the simulation, where the intrinsic timescales can
be much larger, is required to integrate with unnecessarily short
time-steps, leading to a large computational cost.  However, the
multipatch method, in which different regions are updated by independent
processes, allows each patch to have its own time-step while
nonetheless evolving the system in a fashion synchronized across all
patches.  We call this mode of operation ``heterogeneous
time-steps", in contrast to the simpler ``homogeneous time-step" mode
in which all patches are forced to have the same time-step.

Heterogeneous time-steps can be managed with great flexibility. The
only restriction placed on the time-steps in different
patches is that the update times should all be synchronized at intervals
equal to the longest of the time-steps, $\Delta T \equiv \max_k (\Delta t_k)$,
where, as before, $k$ is an index labeling the different patches.  To optimize
computational resource use, before initiating a run the user adjusts
the number of CPUs assigned to each patch so that the wall-clock time to advance by 
a time $\Delta T$ is approximately the same for all patches.

In practice, the coordination works as follows.  At the $n$th synchronization
time $t_n^{\rm sync}$, the different patches exchange boundary condition data.
They then also exchange information about their time-steps so they can
determine which is the longest and in which patch it is found (call that
patch $K$). If all the other patches receive their boundary data either
from patch~$K$ or the problem boundary, the next synchronization time is
set to be $t^{\rm sync}_{n+1} = t^{\rm sync}_n +\Delta T$.
If $\Delta T$ is a factor $Q_l$ ($> 1$ by definition) larger than
the time-step $\Delta t_l$ in some other patch~$l$, patch~$l$
performs $\sim [Q_l]$ updates, where $[x]$ is the greatest integer
$\leq x$, while patch~K works to advance to $t^{\rm sync}_{n+1}$.
When patch~$l$ reaches a time $t^\prime$ such that
$t^{\rm sync}_{n+1} - t^\prime < \Delta t_l$, $\Delta t_l$ is reset
to $t^{\rm sync}_{n+1} - t^\prime$ to achieve synchronization.
Patches that have arrived at $t^{\rm sync}_{n+1}$ before the rest of
the patches wait until all have reached it.  When that has been
achieved, the cycle is repeated.  Because, by definition, conditions in
patch~$K$ change by at most a modest amount over a time $\Delta T$,
it is unnecessary for the other patches to receive boundary condition
information from it during their individual time-steps within the interval
$\Delta T$.  However, when there are more than two patches, it is
possible that some patches may require boundary data from other
patches whose time-steps are shorter than $\Delta T$.  When that
occurs, that pair must exchange boundary data at times determined
by the longer of their two time-steps.  Note that processors {\it within}
the same patch trade boundary data in the usual way at each of that
patch's internal time-steps.

If all the patches are solving the same equations, optimal load balancing
can be achieved when patch~$K$ can be identified with reasonable reliability
in advance, and the ratios $Q_l$ can similarly be estimated.  If those
criteria are met, all that is necessary is to assign processors in patch~$l$
a number of cells $N_l \simeq N_K/Q_l$.  Depending on system architecture, this simple
load-balancing method may be constrained by the total memory available
to processors supporting large numbers of cells.

In some situations, the $Q_l$ might be essentially fixed throughout
the simulation.  For example, this would be the case in a simulation
of gas dynamics in an isotropic gravitational potential in which the
patches are nested spherical shells.  In such a situation, the time-step
for each shell would always be $\simeq [(N_{\phi,k}) \Omega(r_{\rm min,k})/2\pi]^{-1}$,
where $N_{\phi,k}$ is the number of azimuthal cells in patch~$k$,
$\Omega(r)$ is the orbital frequency as a function of radius, and
$r_{\rm min,k}$ is the smallest radius in patch~$k$.  In such a case,
load-balancing could be achieved fairly reliably and would need no
adjustment during the simulation.

More often, however, the $Q_l$ may vary as functions of time.
When this condition obtains, because the MPMD environment does not permit
dynamic reassignment of processors from one program to another, perfect
load-balancing through adroit assignment of processors to patches will nearly always be
an unreachable goal.  Nonetheless, as we show in Sec.~\ref{sec:efficiency},
even approximate load-balancing by combining appropriately chosen numbers
of processors per cell in each patch with heterogeneous time-steps can lead to significant gains
in computational efficiency relative to homogeneous time-step operation.
These gains can be sustained even if the ratios
$Q_l$ change significantly through the simulation if the user periodically
stops the simulation and restarts with an adjusted choice in numbers of
processors per patch.

\section{Physics tests}
\label{sec:tests}

In this section, we showcase the performance quality of the multipatch
method. The tests appropriate to this system are different from those
useful to verify fluid codes because the multipatch infrastructure does
not directly update fluid quantities; rather, it transfers results from
one region to an adjacent one.  Consequently, the focus of our tests is
\Patch's ability to bridge patches without undermining the quality
of the underlying code's solution of the fluid problem.

That the issue is enforcing consistency between patches rather than
the quality of the solution within individual patches explains why we
do not present special tests of the method's multiphysics capability.
Nearly all examples of local physics (e.g., viscosity, different
equations of state) affect the way the fluid state variables (mass
density, internal energy density, velocity/momentum density) behave,
but do not change which variables are transmitted from patch to patch.
Consequently, if our system works for a homogeneous physics example,
it works just as well for a multiphysics example.  Even if magnetic
fields are important to the physics in one patch, but not others, it
must be the case that for some reason (e.g., high resistivity)
magnetic fields weaken greatly near the borders of that single patch.
There is then no need to transfer magnetic field data to the other
patches because it is not relevant to them.

First we demonstrate that it accurately reproduces the analytic
solutions to two classic hydrodynamic simulation test-cases even
when critical features of these solutions pass through patch boundaries
and the grid symmetries and resolutions of the patches differ sharply.
We then explore how well non-conservative interpolation
maintains conservation of mass, momentum, and energy, and identify the
conditions in which it does not.  

\subsection{Sod shock tube}

In this test, we demonstrate that patch boundaries create no
significant artifacts when shocks and rarefaction waves travel from one patch to
another.  For this test, we created a square planar problem
volume in which, following the Sod prescription \citep{Sod1978}, the
fluid is initially at rest everywhere, but there is a sharp pressure
and density discontinuity at a specific value of $x$ within the
volume.  There is no initial variation as a function of $y$.  Within
this volume, we placed a local patch and gave it a constant velocity
so that it moves diagonally in the $xy$-direction.  We performed four
runs to demonstrate the code's performance in a variety of coordinate
system configurations.  In all, the background spacetime is taken to
be Minkowski.

In three of these, the coordinates for both the global and local patches
are Cartesian, while the fourth uses Cartesian coordinates for the global
patch and cylindrical for the local patch.   All Cartesian-Cartesian configurations
have aligned local and global grids, and in all three the global patch cells are
$8\times$ the size of the local patch cells. Because the cylindrical-Cartesian test
is designed to explore sensitivity to grid symmetry contrast rather than gridscale
contrast, the uniform cylindrical grid of the local patch has cells comparable
in size to the Cartesian cells of the global patch.

In two of the Cartesian-Cartesian tests, the local patch moves slowly
relative to the global patch; in one of these tests, the local patch
is placed so that the shock passes through it, while in the other the
local patch is placed where the rarefaction wave runs across it.  In
the third Cartesian-Cartesian test, the discontinuity runs through the
global patch at the start, but the local patch travels rapidly enough
to run through the rarefaction wave, the contact discontinuity, and the shock
 and then emerge on the far side.  The problem solved in the
cylindrical-Cartesian test is similar to the third
Cartesian-Cartesian test in that the discontinuity starts within the local patch,
while the local cylindrical patch moves at the slower
velocity used in the first two Cartesian-Cartesian tests.

Because \HARM is framed in terms of relativistic dynamics, it is convenient
to choose $c$ as the unit of speed.  Given arbitrary code-units of length $\ell_0$
and mass density $\rho_0$, the unit of time is $\ell_0$ and the unit of
pressure is $\rho_0 c^2$.  To ensure Newtonian flow \citep{Hawley1984},
it suffices to make $p/\rho \ll 1$ when measured in code-units.  We also
chose an adiabatic index $\gamma=1.4$. 

Our problem volume was 40 code-units on a side.  For the Cartesian-Cartesian
tests, there were $400^2$ cells in the global patch, each with dimensions $0.1\times 0.1$.
The three Cartesian local patches had side-lengths of 8 and were cut into
$640^2$ cells of dimension $0.0125\times 0.0125$, so that each cell was 1/8 the
size (per dimension) as those in the global patch.  For the cylindrical-Cartesian
test, the global patch had $800^2$ cells, each $0.05 \times 0.05$, while the
local patch's cylindrical grid consisted of $240$ uniform cells over $2 < r < 8$ in
cylindrical radius and $1000$ uniform cells over the full azimuthal extent.
The cylindrical local patch requires a cut-out at its center so as
to avoid the coordinate singularity at the origin associated with polar
coordinates. In this cylindrical grid, the largest azimuthal cell size was
approximately equal to the global patch's cell size while the radial cell-width
was half the global patch's cell size.

  For the initial state, the gas was divided into left
(L) and right (R) states, with density and pressure $\rho_L = 1.0
\times 10^5$, $p_L = 1.0$ and $\rho_R = 1.25 \times 10^4$, $p_R =
0.1$.  The sound speed was therefore $\simeq 3$--$4 \times 10^{-3}$ on
both sides, clearly sub-relativistic.  Zero-gradient boundary
conditions were used for the problem exterior.  In the ``shocked slow patch"
test, the state divide was placed at $x=-6$, while it was located
at $x=0$ for the cylindrical-Cartesian case, at $x=7$
for the ``fast patch" run, and at $x=6$
for the ``rarefaction slow patch" case.
The local patch's origin  initially coincided with the  point $(-15,15)$ in the fast patch simulation,
while all other Sod tests began with the local patch centered at the point $(0,10)$.  
When the patch moved slowly, its velocity was
$\vec{V} = 10^{-3} (\hat{x} - \hat{y}) / \sqrt{2}$; i.e. it traveled
subsonically and considerably slower than the shock front. When the local
patch moved fast, its velocity was $\vec{V} = 5 \times 10^{-2} (\hat{x} - \hat{y}) / \sqrt{2}$,
i.e. it traveled supersonically and $\simeq 6.2\times$ faster in the $\hat{x}$-direction than the shock front. 

\begin{figure*}[ht!]
\figurenum{4}
\plotone{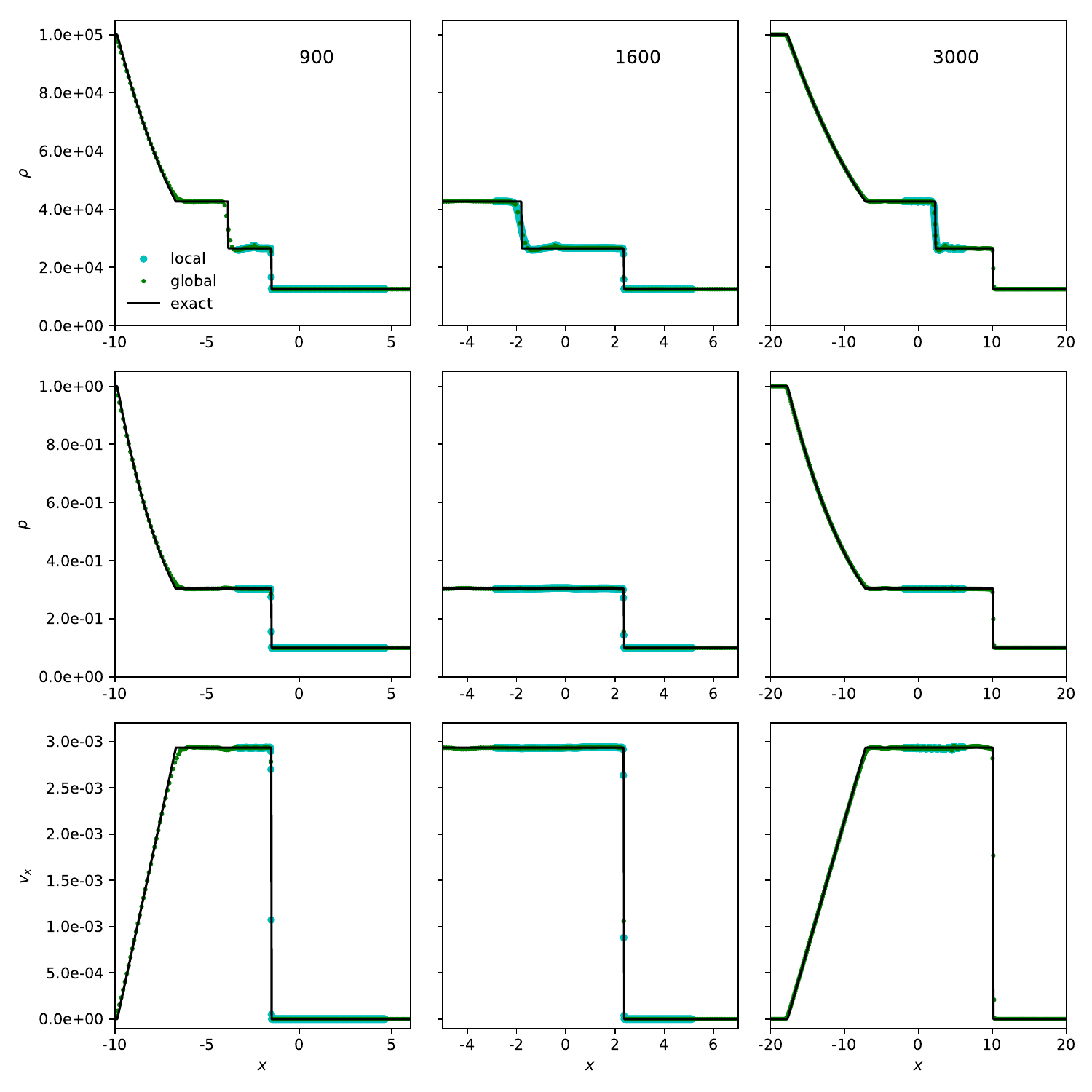}
\caption{Shock tube test problem for the Cartesian-Cartesian ``shocked slow patch"
case, observed at three times, $t=900$ (left), $t=1600$ (middle), and $t=3000$ (right).
Note that the horizontal axis scales are different for all three times in order to highlight
different segments of the problem.
  Each column of three panels shows 1D cuts in density $\rho$, pressure $p$,
  and velocity $v_x$ as functions of $x$ at $y=10$.  Data from the global
  patch is shown with small green dots, data from the local patch with large
  cyan dots, data from the analytic solution is shown with a black line.
  }
\label{fig:sod-slow-left}
\end{figure*}

The results of all four cases can be compared with exact analytic
solutions \citep{Laney1998}.  In Figure~\ref{fig:sod-slow-left}, we
show data from the ``shocked slow patch" run.  At $t=900$ (the left
column), the shock has just entered the local patch; at $t=1600$
(middle column), both the shock and the contact discontinuity run
through the local patch; at $t=3000$ (right column), the shock has
exited the local patch, but the contact discontinuity remains within
it.  In all stages, the multipatch solution follows closely the exact
analytic solution.  The only noticeable departure is a slight
smoothing of the contact discontinuity, visible in the density plot at
$t=1600$, due to the fact that the discontinuity formed in the coarser
global patch.  Although almost invisible in these plots, it is also
worth pointing out an obvious consequence of the multipatch approach:
the shock wave is always only two or three cells thick, and is
therefore considerably sharper in physical space where it runs through
the local patch.  We note that where the global and local patches
overlap, the global patch values plotted are those interpolated from
local patch values even though the values may not be used in the
global patch's evolution; all other global patch values shown are
those resulting from the global patch's update procedure.

\begin{figure*}[ht!]
\figurenum{5}
\plotone{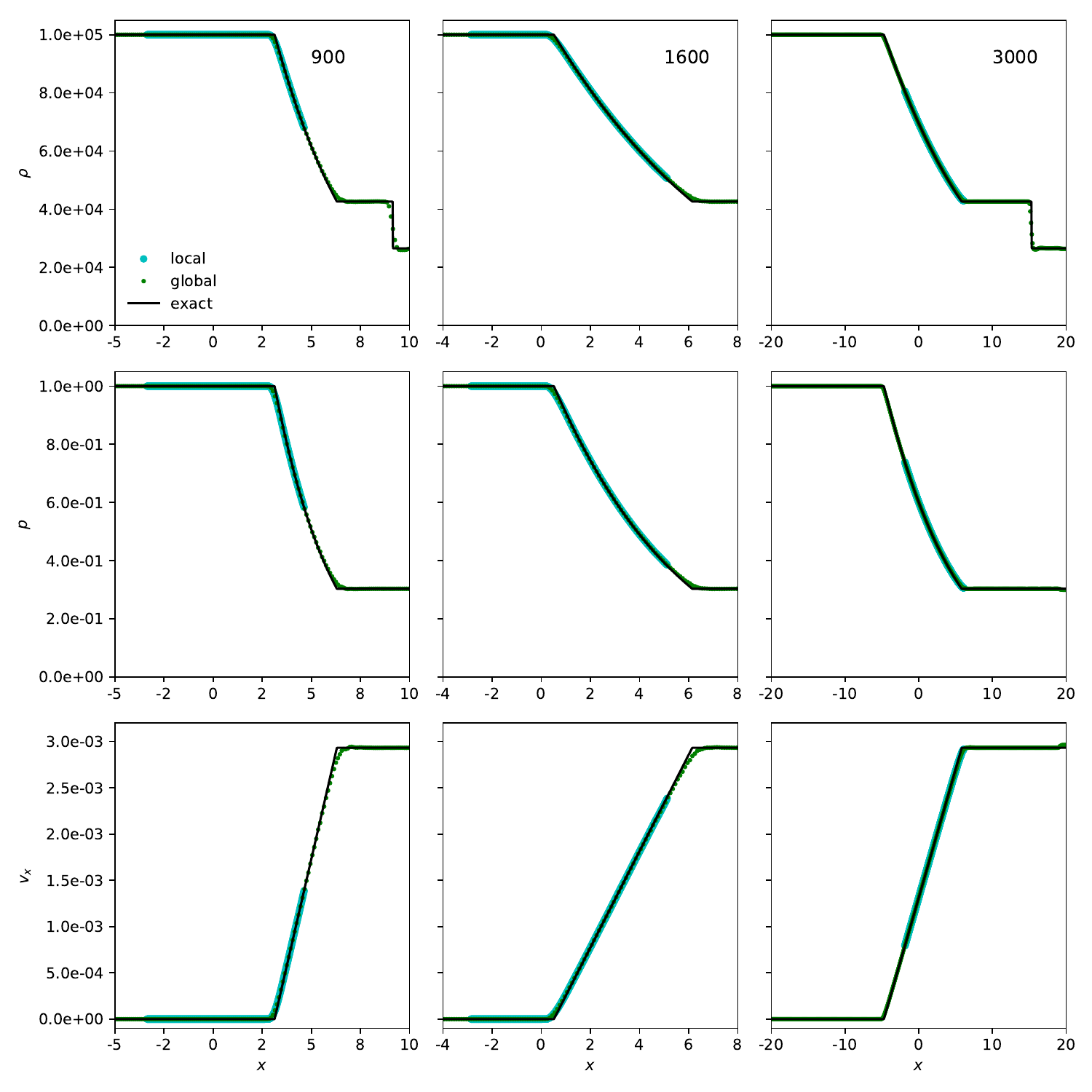}
\caption{Shock tube test problem for the Cartesian-Cartesian ``rarefaction slow patch"
case, observed at the same three times as in Fig.~\ref{fig:sod-slow-left}, $t=900$ (left),
$t=1600$ (middle), and $t=3000$ (right).  Again, the horizontal axis scales are different
for all three times in order to highlight different segments of the problem, and
  each column of three panels shows 1D cuts in density $\rho$, pressure $p$,
  and velocity $v_x$ as functions of $x$ at $y=10$.  Symbols are also as in Fig.~\ref{fig:sod-slow-left}.
  }
\label{fig:sod-slow-right}
\end{figure*}

The second test, the ``rarefaction slow" case, is shown in Figure~\ref{fig:sod-slow-right}.
In all three snapshots, part, but not all, of the rarefaction wave is contained in the local patch.
Where the multipatch formalism has affected the results, agreement with the analytic
solution is essentially perfect; the only departures are a very slight rounding of the
trailing edge of the rarefaction wave apparent at the earlier times when the local patch
has never been anywhere near this edge.  These departures are the same size as those seen
in when using a single patch at the global patch's resolution. 

\begin{figure*}[ht!]
\figurenum{6}
\plotone{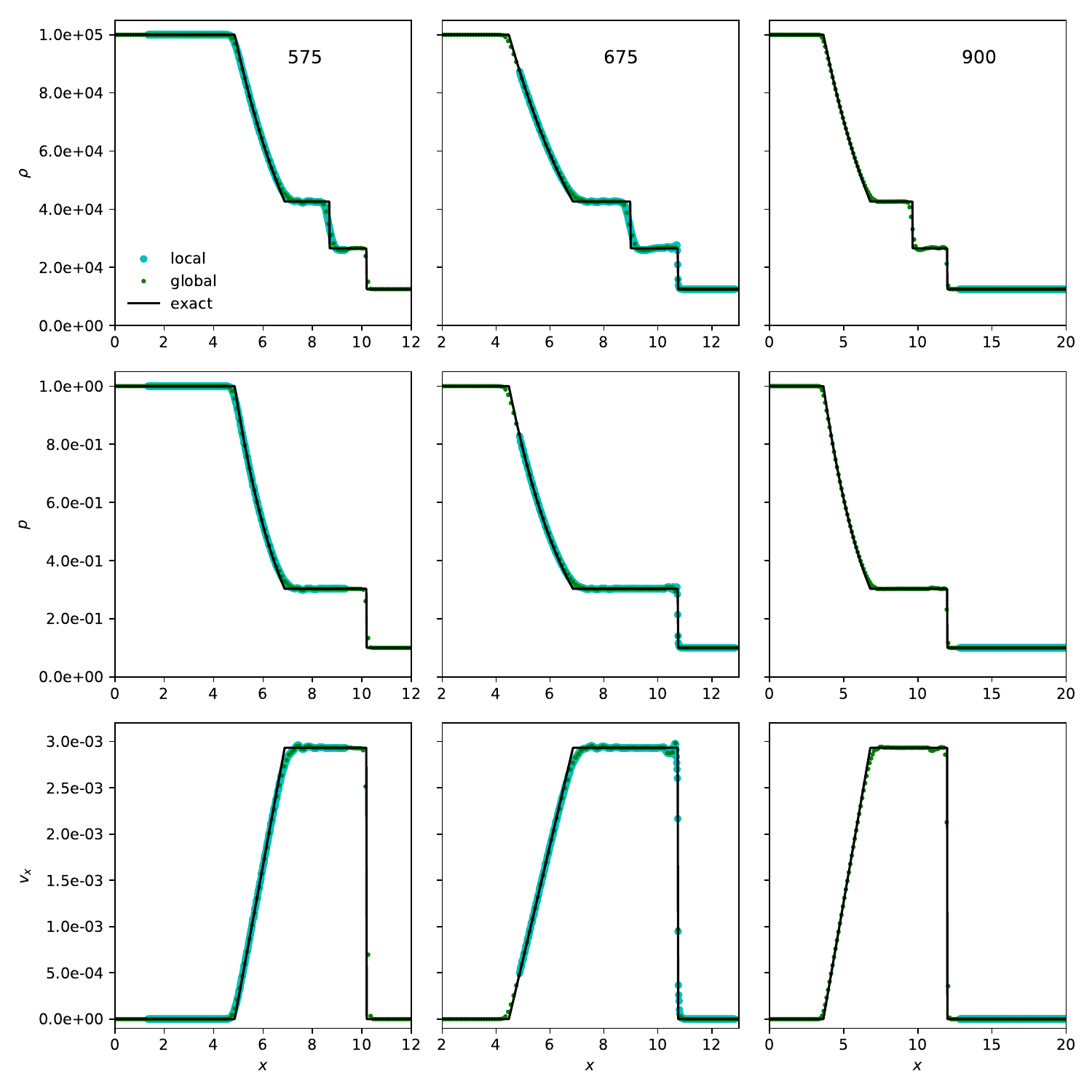}
\caption{ Shock tube test problem for the Cartesian-Cartesian ``fast patch"
    case, observed at three times, $t=575$ (left), $t=675$ (middle), and $t=900$ (right).
 Again, the horizontal axis scales are different
for all three times in order to highlight different segments of the problem, and
  each column of three panels shows 1D cuts in density $\rho$, pressure $p$,
  and velocity $v_x$ as functions of $x$ at $y=3$.
  Symbols are also as in Fig.~\ref{fig:sod-slow-left}.
  }
\label{fig:sod-fast-cart}
\end{figure*}

The third Cartesian-Cartesian run tests whether supersonically
  moving local patches create any special problems.  The results of
  this test are shown in Figure~\ref{fig:sod-fast-cart}.  As can
  be seen, small errors are induced in the fluid velocity between
  the shock front and the rarefaction wave.  However, these are
  {\it not} due to any property of the multipatch method: similar
  errors are produced in conventional monopatch calculations whenever
  the Mach number of the reference frame's velocity relative to the shock
  velocity is significant.  We have compared the errors seen in this
  multipatch test to those seen in a simulation of the identical problem
  in which only the global patch is present, i.e., a conventional
  monopatch run, but with the gas initially given a bulk uniform velocity.
  In this comparison run, the errors are $\simeq 2\%$.
  The errors in the multipatch test are, depending on
  location, generally smaller than, but in a few places comparable to, those in the
  monopatch run.  We have also repeated
  the multipatch test illustrated in Figure~\ref{fig:sod-fast-cart}
  with a less extreme local patch velocity, a $\hat{x}$ velocity relative to the
  shock front only $2 \times$ the shock speed rather than $6.2$.
  The errors in this test are also typically 
  smaller than, but in a few places comparable to, those in the figure. 
  We therefore expect at most modest-amplitude errors when high-contrast
  local patches pass at high Mach numbers through shocks, and considerably
  smaller errors when the relative speed is small, as would often be
  a desirable choice.  Indeed, one of the advantages of the multipatch
  method is the flexibility it offers to choose preferred reference frames
  in different portions of the problem.

 In our last Sod test case, the cylindrical-Cartesian run, the
  shock traverses a different sequence of patches as a function of the 
  $y$-coordinate.  For instance, the shock wave traveling
  along $y=10$ from $x=0$ starts in the global patch, then enters the
  local patch through its inner radial boundary, and ultimately exits the
  local patch's outer radial boundary as it re-enters the global patch.
  Along other constant-$y$ trajectories, the waves may start in the local
  patch and emerge in the global patch, or always reside in the global
  patch.

\begin{figure*}[ht!]
  \figurenum{7}
  \begin{center}
  \begin{tabular}{rl}
    \multirow{4}{*}{\includegraphics[width=0.4\textwidth]{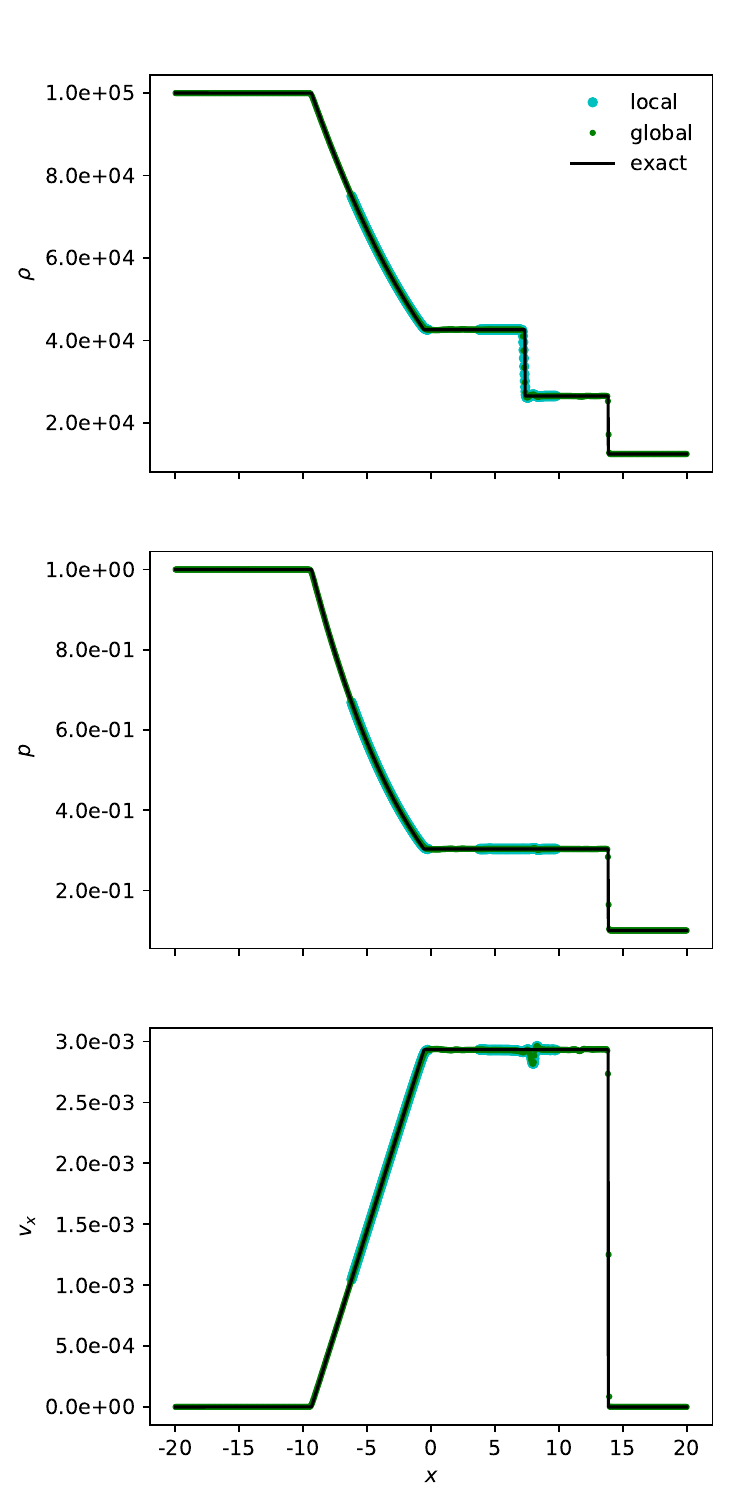}}\\
   &     \includegraphics[width=0.4\textwidth]{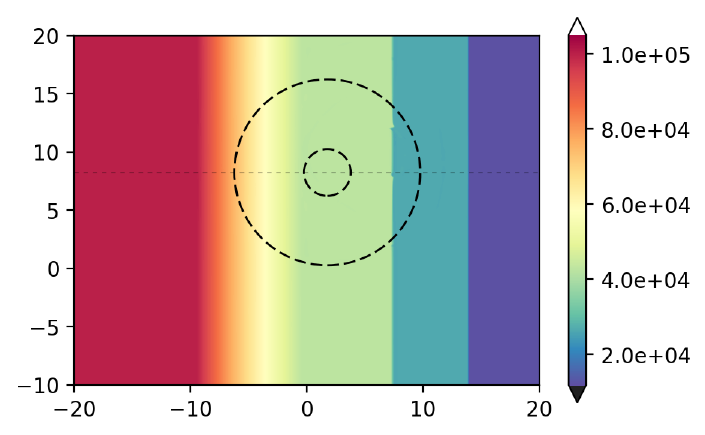}\\
   &     \includegraphics[width=0.4\textwidth]{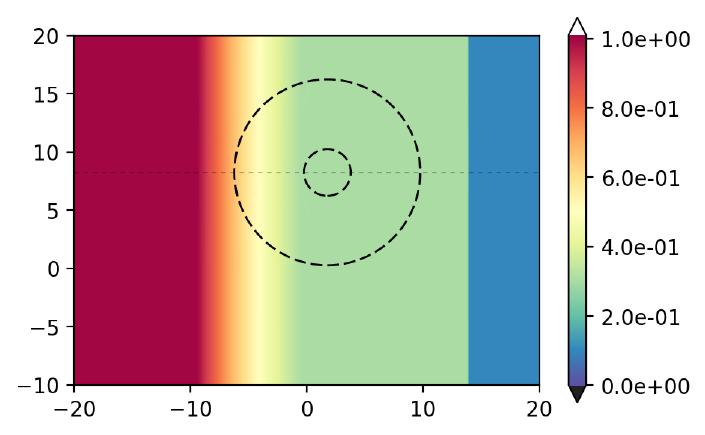}\\
   &     \includegraphics[width=0.4\textwidth]{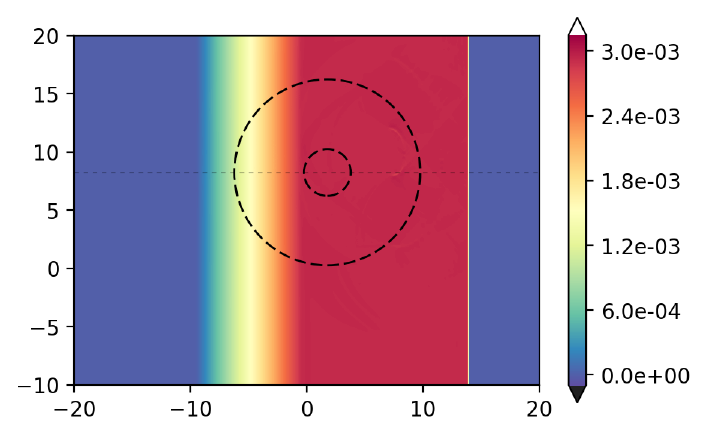}\\
   \end{tabular}
  \end{center}
  \caption{Shock tube test problem  in which the local patch used
      cylindrical coordinates and the global patch used Cartesian
      coordinates, observed at $t=2500$.  The rest-mass density
    ($\rho$, top row), pressure ($p$, middle row), and $x$-component
    of the velocity ($v_x$, bottom row) are shown.  Their full 2D
    contours (right column) are shown next to slices (left column)
    taken along $y=8.23$, where departures from the exact solution are
    the largest.  In the 2D plots, the local patch data (outlined in
    dark long dashes) are shown on top of the global patch data, and
    the location of the slices is displayed (light, short dashes).
    The line plots use the same conventions used in
    Figure~\ref{fig:sod-slow-left}.  }
\label{fig:sod-cyl}
\end{figure*}

 This test's results are illustrated quantitatively in
 Figure~\ref{fig:sod-cyl}.  Unlike the situation in the first three
 tests, the local patch's grid here no longer conforms to the symmetry
 of the initial data.  We show 2D contours of the three fluid
 quantities from the run to demonstrate how well \Patch maintains the
 problem's linear symmetry despite the cylindrical local patch.  While
 the rarefaction and post-shock states show no signs of $y$-asymmetry,
 we do find minor artifacts at the contact discontinuity (at $x \simeq
 8$ in the images) in $\rho$ and $v_x$. These artifacts are so small
 they are difficult to see except in the line plot of $v_x$.  The
 artifacts in $\rho$ take the form of a one or two cell displacement
 of the contact discontinuity. They originate from the two points
 where the newly-formed contact discontinuity intersects the local
 patch's inner boundary and then travel with the contact
 discontinuity.  The artifact in $v_x$ also begins when the contact
 discontinuity crosses the local patch's inner boundary and similarly
 travels with the contact discontinuity.  It, however, takes the form
 of a $\sim 5\%$ error in $v_x$ along a half-circle ``echo'' of the
 patch boundary.  The amplitude of the artifacts decreases with finer
 grids. Also, the artifacts are not cylindrically symmetric on the
 local patch because they are advected with the $x$-oriented velocity
 of the solution while the local patch moves diagonally in the $x-y$
 plane.

\subsection{Sedov-Taylor blast wave}

The purpose of this test was to show the performance of the multipatch
when at least one of the patches has a grid whose symmetry is a poor match
to the natural symmetry of the problem  and to demonstrate that crossing a
patch boundary separating regions of different grid symmetry introduces no ill
effects.  To that end, we study a Sedov-Taylor
3D spherical blast wave \citep{Sedov1959} with a central local patch using
Cartesian coordinates and a global patch using spherical coordinates.  As
a standard of comparison, we also contrast a monopatch simulation with entirely
Cartesian coordinates.   Although the Cartesian grids are
poor matches to the spherical symmetry of the physical problem, they do have
the virtue of  eliminating the coordinate singularity at the origin created
by spherical coordinates.

A blast wave is formed when a large amount of energy $E$ is deposited in a small
region.  If the ambient gas is motionless, a spherical
shock wave travels rapidly outward.  Once the mass swept up by the shock exceeds
the mass originally located in the small energy-deposition region, the shock
front's radial position as a function of time is given by
\be \label{eqn:Sedov}
R_s = \left( \xi \frac{E}{\rho} \right )^{1/5} t^{2/5}
\ee
until $E/R_s^3$ is small enough to be comparable with the ambient pressure.
Here $\rho$ is the initial (uniform) density of external gas, and
$\xi$ is a dimensionless number $\sim 1$.

To simulate this, we follow \citet{Fryxell2000} and divide the initial state
into two regions.  As in the Sod shock tube problem, we choose the unit of
velocity to be $c$, but use arbitrary code-units for length and mass.
In terms of these units, region~1 is a small sphere of radius $\delta r=25$.
Its initial pressure $p_1 = (\gamma-1) E / (4\pi \delta r^3) = 1$ for adiabatic
index $\gamma$ (again $=1.4$), while its density $\rho_1 = 10^{-3}$.  Region~2
is everything outside $r=\delta r$.  Here the initial pressure $p_2 = 10^{-10}$
and initial density $\rho_2 = \rho_1$.  The dimensionless coefficient
of eqn.~\ref{eqn:Sedov} is a function of $\gamma$; for $\gamma = 1.4$, it is
1.175 \citep{Ostriker1988}. 

For the monopatch, the computational domain is a cube of dimension
2500 having $N_{\rm mono} = 400^3$ equal-volume cubical zones
with side-length 6.25.  We perform two multipatch simulations
in order to illustrate its dependence on gridscale contrast.  In both,
the local patch is a cube of side-length
600 centered on the origin with $120^3$ equal-volume cubical cells of
side-length 5, similar to the Cartesian cell-size in the monopatch simulation.
Likewise in both simulations, the global patch is
a sphere of radius 1000 described in spherical coordinates, but with
two cut-outs: a sphere of radius 290 surrounding the origin
and a bi-cone of half opening-angle $\pi/10$ surrounding the polar axis.
The two multipatch simulations differ in global patch resolution.
In the ``low contrast" case, the angular grid is uniform, with 120 cells
in polar angle $\theta$ and 320 cells in azimuthal angle $\phi$, but the
radial grid has 80 logarithmically-spaced cells.  In this case,
the radial cell size in the global patch at the patch boundary ($r \simeq 300$--400)
is similar to the local patch cell size.  In the ``high contrast"
case, the cell counts in all three dimensions are a factor of 4 smaller, so
that radial cells at the patch boundary are separated by $\simeq 33$,
roughly a factor of 4 larger than local patch cells.

\begin{figure*}[ht!]
\figurenum{8}
\plotone{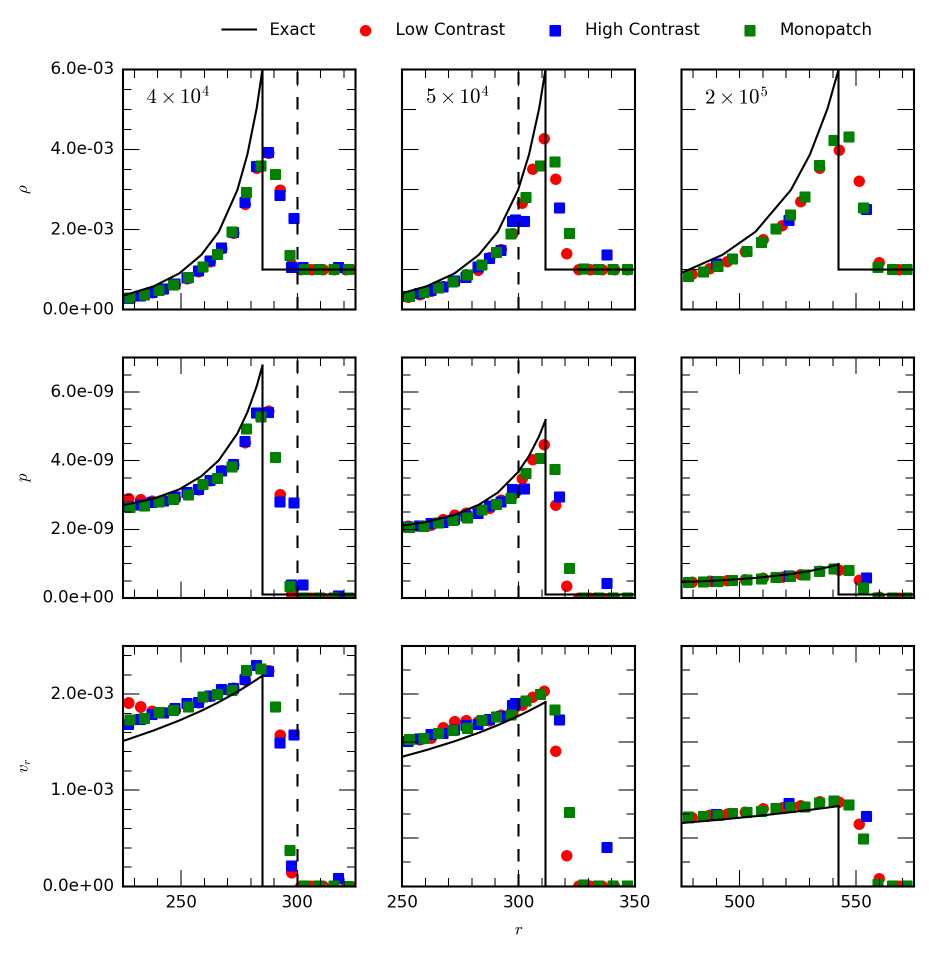}
\caption{ 
\label{fig:sedov}
Sedov-Taylor 3D spherical blast wave at three different times ($t = 4\times 10^4,
5\times 10^4, 2 \times 10^5$).   Monopatch data (green squares) are
contrasted with multipatch data from two different simulations (red
circles for the ``low contrast" case, a global patch with resolution similar to the monopatch,
blue squares for a simulation whose global patch has resolution $4\times$ coarser).
The analytic solution is represented by a black
line.  Each column of three panels shows 1D radial cuts in density $\rho$,
pressure $p$, and radial velocity $v_r$.}
\end{figure*}

We portray how well the multipatch simulations do, relative to
both the analytic solution and the monopatch simulation, in 
Figure~\ref{fig:sedov}, which again shows the situation at three different
times.  At the earliest time, the shock front is entirely within the local patch, while it is
a short distance outside the local patch in the middle time, and far
outside the local patch at the last time.  At the earliest time, the
data for the Cartesian local patch and the Cartesian monopatch are, not
surprisingly, nearly identical; the entire global patch remains in the
initial state at this time.  Interestingly, the shock at this time is
at slightly larger radius than predicted by the analytic solution
in both the monopatch and the multipatch simulations.

At the middle time, the local patch and monopatch still closely agree, but the
shock region is located in the global patch.  As is clear from the
curve showing the analytic solution, when the shock has left the
local patch, a gridscale $\lesssim 10$ is a prerequisite for describing
the density and pressure profiles.  The low contrast multipatch case therefore
does reasonably well, slightly out-performing the monopatch; where
the high contrast multipatch case samples the profile, it is in general in good
agreement with both the monopatch and low contrast multipatch data,
but its sampling is too sparse to resolve the actual profile.
This pattern persists to late times: monopatch and low contrast
multipatch behave very similarly to one another; high contrast
multipatch points are placed too sparsely to resolve the profiles,
and their error levels are a bit greater than for the other two
simulations.

Our conclusion from this comparison is that the multipatch method
performs very similarly to a monopatch method.  The poorer performance
of the high contrast case is due entirely to its overly-coarse grid,
a failing that would have very much the same effect if this grid had
been used in a conventional monopatch simulation.

\begin{figure*}[ht!]
\figurenum{9}
\plotone{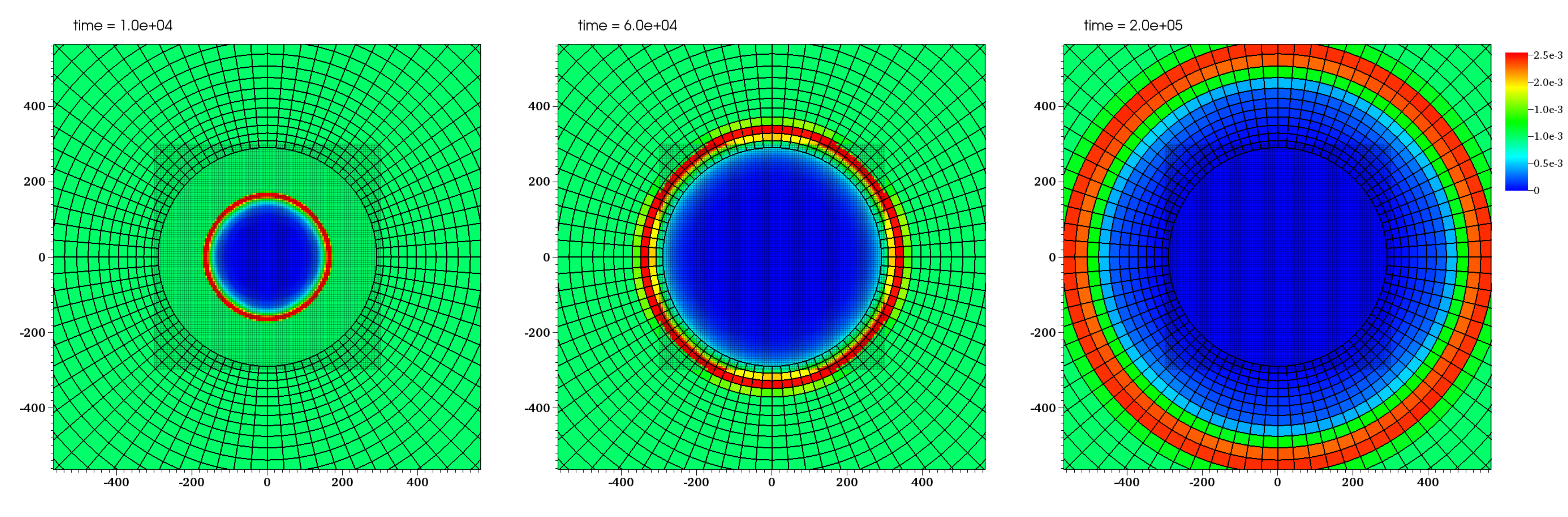}
\caption{ 
\label{fig:SedovImage}
Density in the equatorial plane of the spherical coordinates for
the ``high contrast" multipatch simulation of a Sedov-Taylor 3D spherical
blast wave.  The three panels correspond to the same three times shown in
Fig.~\ref{fig:sedov}.}
\end{figure*}

 Results from an absolute test may be seen in Figure~\ref{fig:SedovImage}.
Here we show how well the high contrast multipatch simulation is able
to support the intrinsic spherical symmetry of the physical problem.  When
the shock still lies within the Cartesian local patch (left panel), its outline
is very nearly circular, but there are small departures from perfect azimuthal
symmetry due to the underlying Cartesian grid.  When the shock lies partly
in the spherical global patch and partly in the Cartesian local patch (middle
panel), the shock is almost perfectly azimuthally symmetric in the global
patch, but where it passes through the local patch retains the same level of
small-scale noise as at the earlier time.  At late time, when the shock
is entirely within the global patch, it shows a very high degree of
azimuthal symmetry.   Thus, in these tests the multipatch system induces no departures
from the true geometric symmetry; such small errors as exist are due
entirely to the symmetry of the grid.

\subsection{Inter-patch conservation }
\label{subsec:conservation}

As remarked earlier, our interpolation scheme is not strictly conservative,
even though many hydrodynamics codes that can be used in concert with our
multipatch method are.  That contrast makes it worthwhile to examine how
large an error may be induced by non-conservative interpolation, and how
that error depends on the interaction between problem character and details
of multipatch implementation.

In principle, this error could depend on many variables.  To simplify the
discussion and focus on what we believe is the principal issue, we study
an idealized problem, one in which matter flows from a patch in which it
has acquired order-unity amplitude fine-scale structure into a more
coarsely-resolved patch.  The parameter that appears to affect conservation
errors the most is the ratio between the lengthscale of the structure and
the resolution scale of the coarser patch.

To illustrate this dependence, we construct a 3D system in which the problem
volume extends from $x=-30$ to $x=+50$ in Cartesian coordinates, but in the
transverse directions ($y$ and $z$) spans only the range $[-20,+20]$.  The
local patch is stationary and occupies the region $-30 \leq x \leq 0$ in global
coordinates.  Both patches have uniform cubical grids that are parallel to each other,
but the cell-sizes of the global patch (5) are $10\times$ that
of the local patch (0.5).

At $t=0$, all of the fluid is traveling at $V_x = 0.1$ in the x-direction,
but its density and pressure differ sharply across the line $x=+10$, located
a short distance into the global patch from the local patch boundary. To
the left of that line, $\rho_L = 1 + \sin^2(\omega_n y)$ and $p_L = 10^{-12}\rho_L$
(as in the previous tests, $c=1$ in our units), while on the right
$\rho_R = 10^{-16}$ and $p_R = 10^{-28}$.
The sharp pressure contrast induces a flow from left to right in the frame of
the bulk flow.   Because the sound speed on the
left is so small ($\sim 10^{-6}$), even in 1000 time-units the high pressure
gas expands only a very slight distance to the right in the moving frame of the
fluid.  Thus, the density and pressure modulation across the patch boundary at
$x=0$ is essentially constant throughout the run of the test.

\begin{figure}[ht!]
\figurenum{10}
\epsscale{1.3}
\plotone{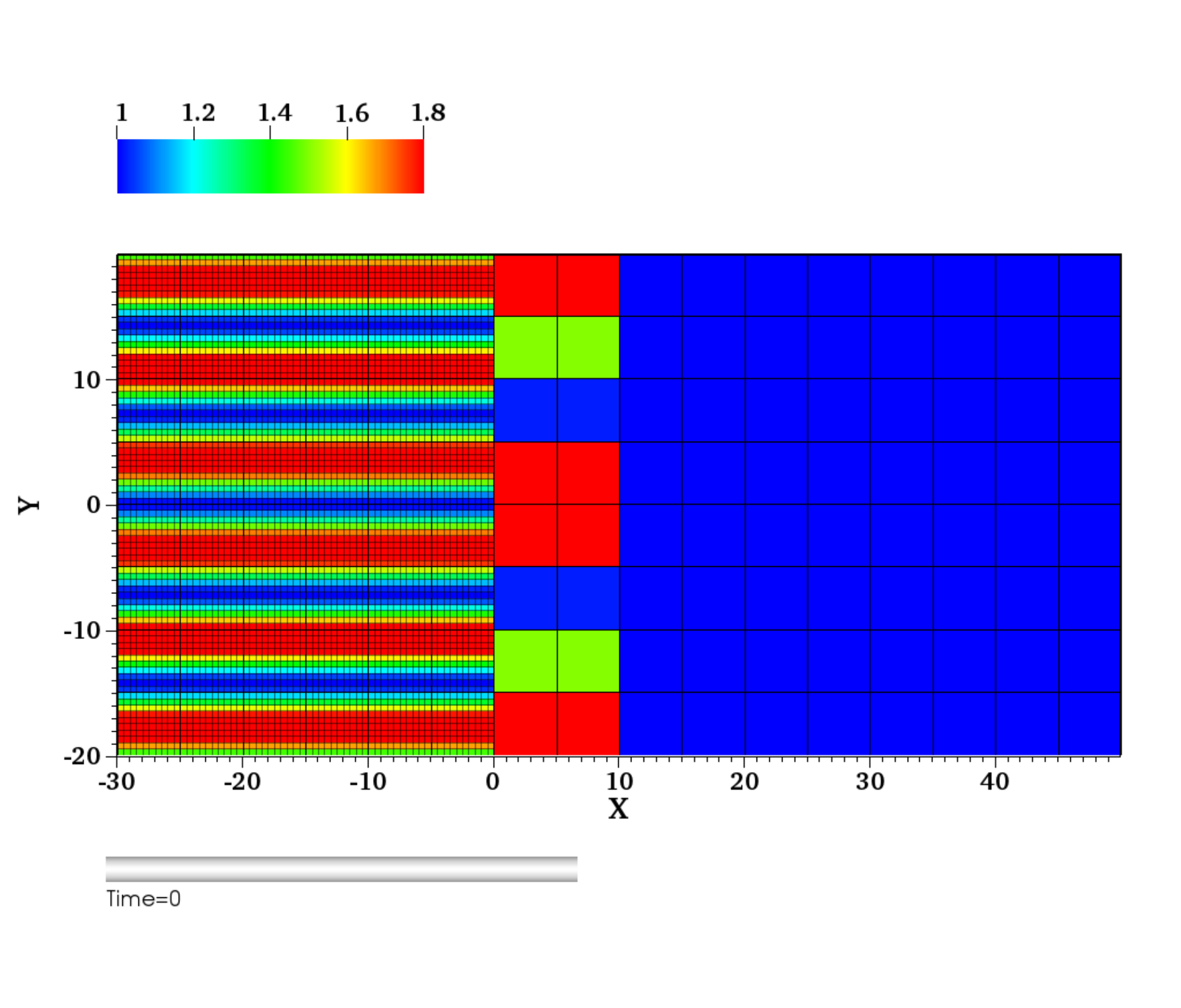}
\caption{ 
\label{fig:interp1}
Density (colorscale) in the initial condition for a conservation test with $n=0.7$.
Global patch grid lines are given in both patches for reference; local patch
grid lines are shown only in the local patch.}
\end{figure}

The frequencies $\omega_n = n \omega_g$ are chosen as multiples of the spatial Nyquist
frequency of the global grid resolution, $\omega_g = \pi/\Delta x$; this definition also
ensures that the total mass in the entire
computational domain is the same for each $\omega_n$. The case illustrated in
Figure~\ref{fig:interp1} is for $n = 0.7$; as can be seen, the coarse grid drastically
smooths the modulation.

Outflow boundary conditions are enforced at $x=50$ while reflecting boundary
conditions are applied at all other physical boundaries.
In the absence of numerical error, these boundary conditions (and the
extremely low value of $\rho_R$ in the initial state) guarantee that
the total mass, energy, and momentum on the grid remain constant until the
front arrives at the far right edge at $t=400$.

\begin{figure}[ht!]
\figurenum{11}
\epsscale{1.3}
\plotone{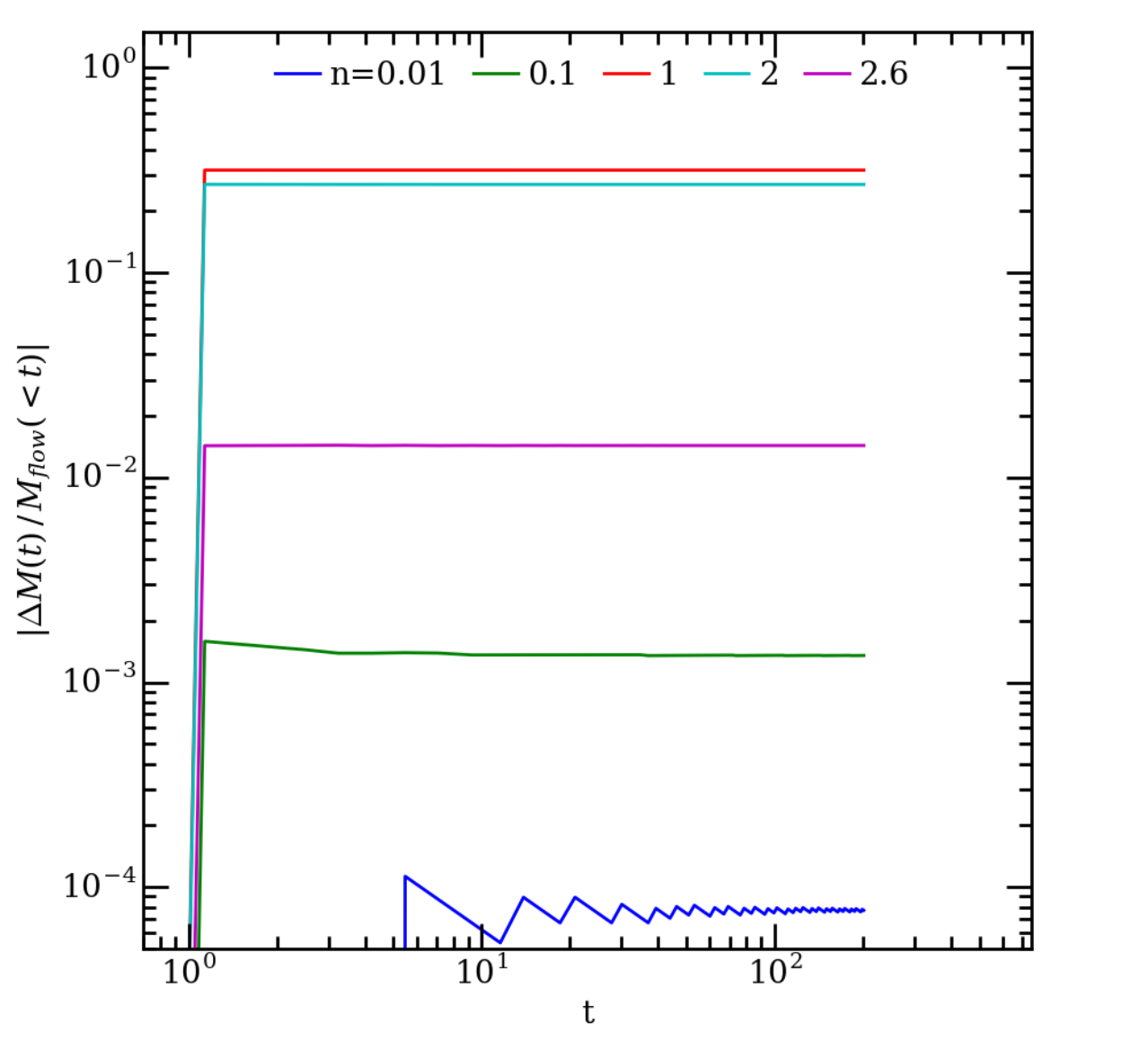}
\caption{ 
\label{fig:interp2}
Change in total mass $\Delta M(t)$ relative to the time-integrated mass flow
across the patch boundary $M_{\rm flow}(<t)$ for initial conditions defined by frequency
$\omega_n$: $n=0.01$ (blue), $n=0.1$ (green), $n=1$ (red), $n=2$ (cyan), and $n=2.6$ (magenta).}
\end{figure}

In Figure~\ref{fig:interp2}, we show the ratio
$\Delta M(t)/M_{\rm flow}(<t)$ between the change in total mass and the amount
of mass that has flowed across the patch boundary up to that time for initial
density modulations with
frequencies $\omega_n = \{ 0, 0.01, 1.0, 2.0, 2.6\} \times \omega_g.$
As is appropriate to the steady-state flow we are studying, the fractional
error is independent of time for all cases.  Also not surprisingly, when $n < 1$,
so that the global grid resolves the modulation well, the error is small:
$\lesssim 10^{-4}$ for $n=0.01$, when there is almost no modulation, and
$\approx 1.5 \times 10^{-3}$ for $n=0.1$.  Once $n \gtrsim 1$, when the global
grid can no longer support the modulation, the pattern changes.  The error
for $n=2.6$ is larger than for $n < 1$, but still tolerable
($\approx 1.5 \times 10^{-2}$).  However,
the error for both $n=1$ and $n=2$ is uncomfortably large: $\approx 0.3$.
These two values of $n$ are special cases: the modulation is resonant with
the global grid pattern, so the error in the mass flow depends strongly
on the phase of the modulation at cell-centers, which is the same for
all global cells.  The value for $n=2.6$ should therefore be more characteristic
of generic modulations.

We have also examined the fractional conservation errors for $x$-momentum and
internal energy, but we do not present them explicitly because the figures are
virtually indistinguishable from those shown in Figure~\ref{fig:interp2}.  That
they should be so similar is to be expected because they are interpolated by
identical procedures.

\begin{figure}[ht!]
\figurenum{12}
\epsscale{1.3}
\plotone{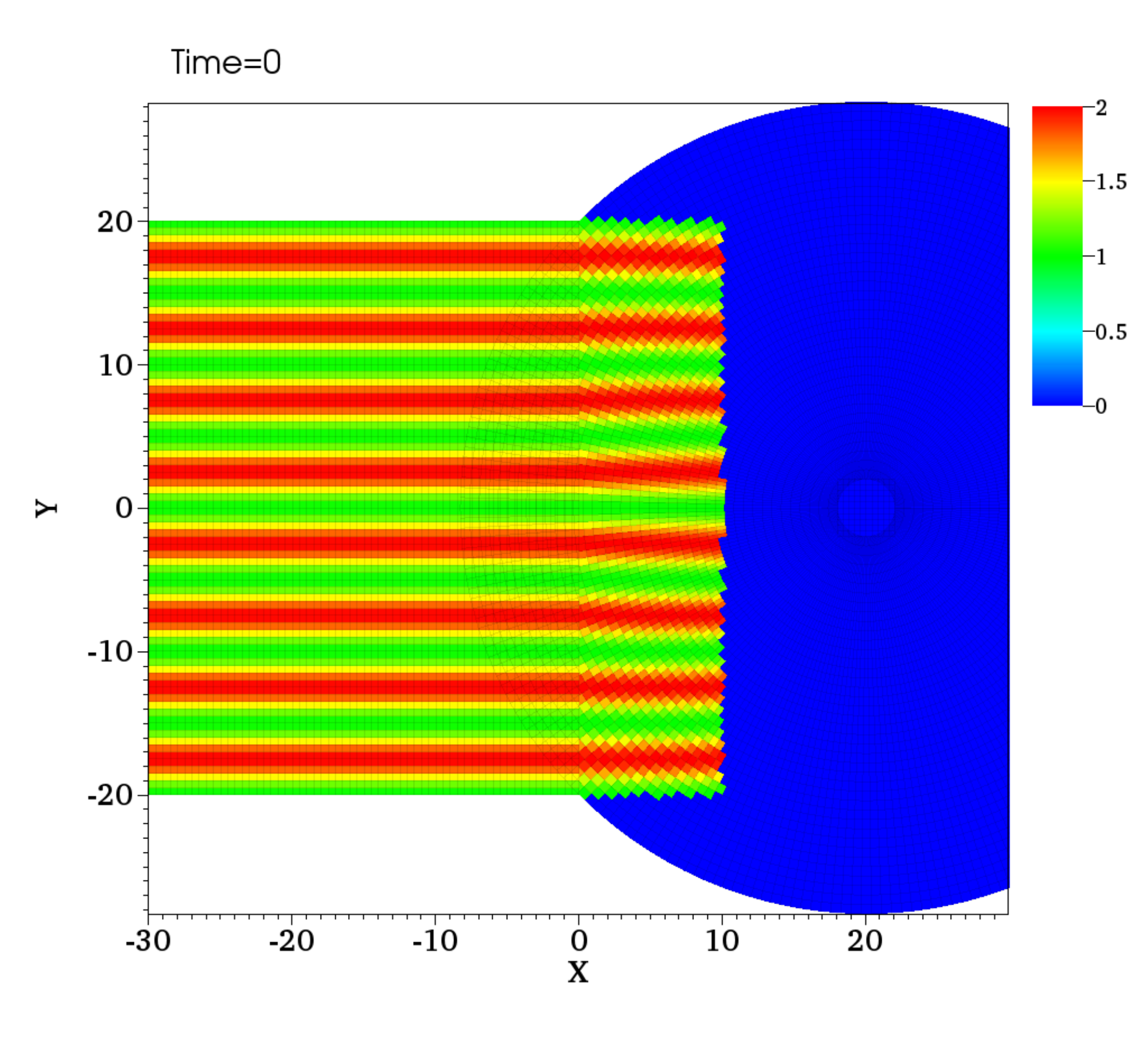}
\caption{ 
\label{fig:interp3}
Initial condition for density in a conservation test with a Cartesian local patch
and a cylindrical global patch.  Here $n=1$.}
\end{figure}

In addition, we have performed simulations of the same problem with a different
global patch, one with polar coordinates.  In this case, we are testing
the robustness of conservation properties with respect to change of grid
symmetry, rather than with respect to change of grid resolution.  For this
reason, the local patch and initial condition structure are identical to
those of the previous conservation test, but the global patch is given
a cylindrical grid (see Fig.~\ref{fig:interp3}).  The origin
of the cylindrical coordinates is placed at $(x,y) = (20.,0.)$; its radial
cells have width 0.5 and its azimuthal cells have width 0.025~radians so
that both cell dimensions near the patch boundary match those of the local patch.
The global grid frequency $\omega_g$ for this test is defined so that it
is equivalent to the one used in the first set of tests, i.e.,
$\omega_g \equiv \pi/(10\Delta x_{\rm local})$.  The polar grid near the
patch boundary should therefore be able to support modulations with values
of $n$ similar to those used before.  On the other hand, even though the azimuthal
cell size becomes even finer near the global patch
origin, the mismatch between polar cell shapes and the rectangular modulation
induces larger errors closer to the global patch origin (also shown in
Fig.~\ref{fig:interp3}).

\begin{figure}[ht!]
\figurenum{13}
\epsscale{1.3}
\plotone{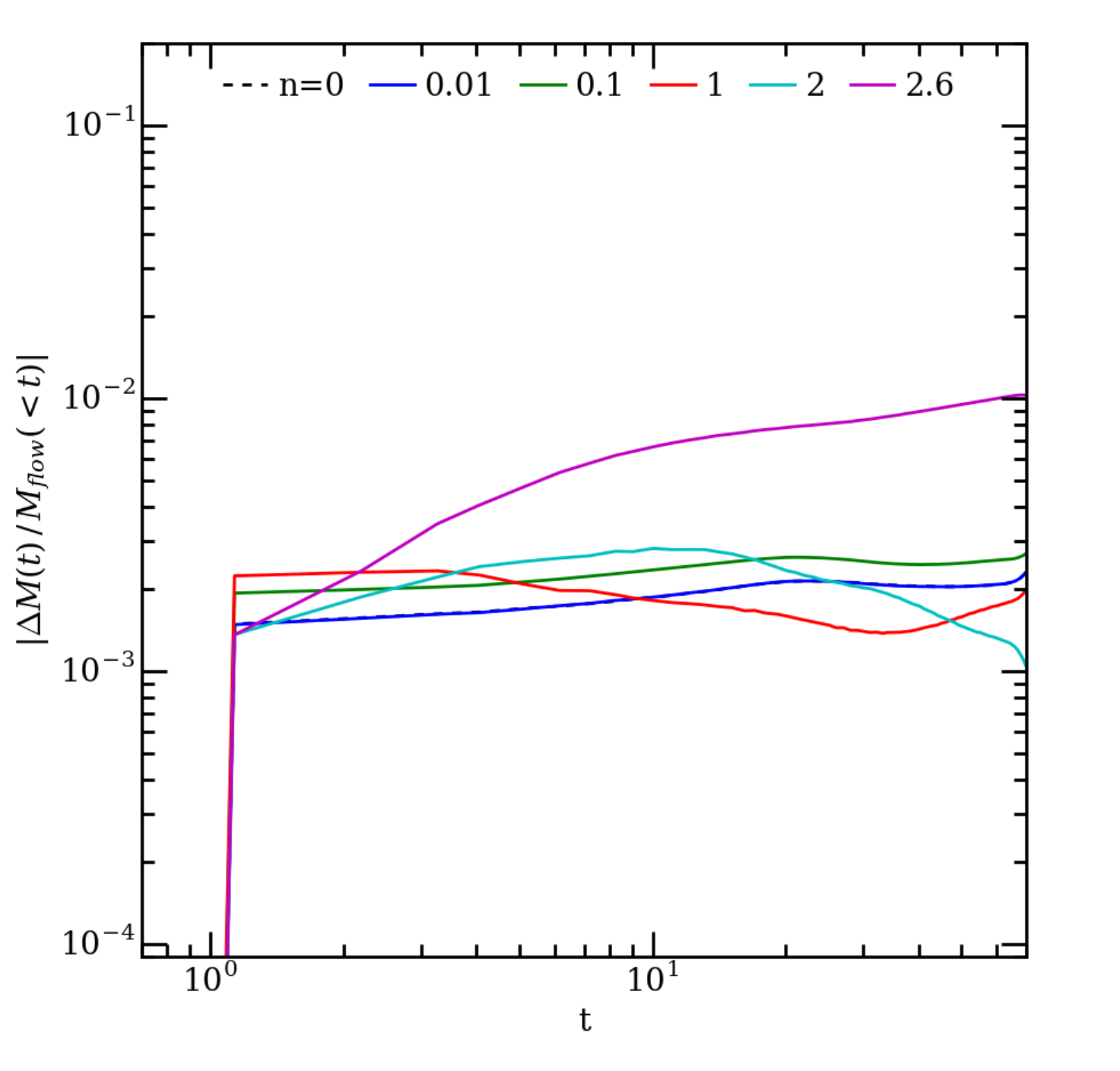}
\caption{ 
\label{fig:interp4}
Change in total mass $\Delta M(t)$ relative to the time-integrated mass flow
across the patch boundary $M_{\rm flow}(<t)$ for initial conditions defined by frequency
$\omega_n$: $n=0.01$ (blue), $n=0.1$ (green), $n=1$ (red), $n=2$ (cyan), and $n=2.6$ (magenta).}
\end{figure}

The results (seen in Fig.~\ref{fig:interp4}) are, nonetheless, comparable to
those from the Cartesian-Cartesian tests.  For all $n \lesssim 2$, the fractional
error is $\approx 2 \times 10^{-3}$, again almost independent of time.  Unlike the
Cartesian-Cartesian case, however, the polar grid eliminates the possibility of resonant
response for integer $n$.  With a polar global grid, the error for $n=2.6$ rises over
time; we believe that this increase is due to the mismatch between the flow properties
and the global grid geometry, a mismatch exacerbated by higher modulation wavenumbers.
If so, it is not a product of errors created as information is transferred across the patch
boundary, but rather one intrinsic to the inappropriate symmetry of the polar grid.
The evidence for this supposition is that if we define the total mass on the grid as
\begin{equation}
M = \int_{-30}^{x_*} \, dx \, \int \, dy \int \, dz \rho,
\end{equation}
we find that the fractional error grows as $x_*$ increases toward $x=20$,
the $x$-coordinate of the polar grid origin.  The values shown in Figure~\ref{fig:interp3}
are for $x_* = 18$.  Again, just as for the Cartesian-Cartesian case,
the error numbers for momentum and energy are virtually identical to those
for mass.

In more realistic problems, we have found that dynamical effects can
enlarge these errors, and in some cases positive feedback loops can
develop.  However, the driving factor appears to be similar, the
inability of a coarse grid to support variations occurring on
lengthscales too small for it to resolve. Devices to curb this sort of
interpolation error are problem dependent.  For example, in our
  tidal disruption test-run (mentioned at the end of
  Sec.~\ref{subsec:multiphysics/scale}), we found that smoothing the
  hydrodynamic variables over the three ghost-cells at the inter-patch
  boundary successfully damped a growing departure from mass
  conservation, holding the error in total mass to $< 1\%$ over a time
  in which the star shed 95\% of its mass.

\section{Computational efficiency and parallelization scaling}
\label{sec:efficiency}

Lastly, we present data on computational efficiency and parallelization
scaling.  When discussing these issues in the context of ordinary monopatch
operation, the principal questions generally have to do with the fundamental
efficiency of the computational algorithm and the ratio between time spent
exchanging boundary condition information and computing updates.  The former
sets the basic scale in terms of zone-cycles per processor per unit time;
the latter is determined by the additional cost incurred by inter-processor
communication.  When a hydrodynamics code parallelizes well, the fraction of total
processor-hours devoted to communication is nearly independent of the
total number of processors.  Thus, to gauge how much overhead is
created by multipatch operation and how efficiently the multipatch system
makes use of parallelization, we must contrast multipatch benchmarks
with monopatch benchmarks treating the same physics problem, and do so
as a function of total number of processors.  In addition, we will explore
how much our heterogeneous time-step option improves efficiency by
contrasting its performance with matched homogeneous time-step runs.

However, all these tests can at best be indicative rather than definitive.
Even in monopatch operation, hydrodynamic code speeds can be problem-dependent,
and there is every reason to expect that multipatch methods will, if anything,
add new ways for code performance to be sensitive to the nature of the
specific problem.  For example, in multipatch problems the amount of computation required to
perform a single zone-cycle can depend on how much effort is necessary
to compute coordinate transformations, a quantity that can easily differ
substantially from one physical situation to another.  If the patches solve
different equations, the number of operations per zone-cycle can change
even more drastically.  Because we expect a significant contrast in overhead
between cases in which the local patches move or are stationary, we
will specifically examine that variety of problem-dependence.

To finesse these complexities as best we can, we focus on a single simple
test problem: evolving a hydrostatic gas in the absence of any external
forces.  The background spacetime is therefore Minkowski, and in the
initial condition there is uniform density, pressure, and entropy. The
fluid's adiabatic index is $\gamma = 5/3$.

The problem volume is a 3D cube treated with two patches, a global patch
and a local patch.  Both use Cartesian coordinates with uniformly-spaced grids.
The local patch is a cube with side-length 1/8 the global patch's side-length,
but has a gridscale that is also 1/8 the global patch's; the two patches
therefore have the same number of cells.  These choices produce a time-step ratio
$\Delta t_g / \Delta t_l = 8$.  These will be studied with two different
numbers of zones per processor, $n = 20^3$ and $n = 40^3$.
Each simulation is run for a fixed time duration, chosen to be just long
enough that initialization time is negligible.  Zero-gradient boundary conditions
are used for the global patch; the local patch never encounters the
problem boundary.  All cases were performed on the same
platform (Texas Advanced Computing Center, Sandy Bridge nodes on Stampede).

\begin{figure*}[ht!]
\figurenum{14}
\plotone{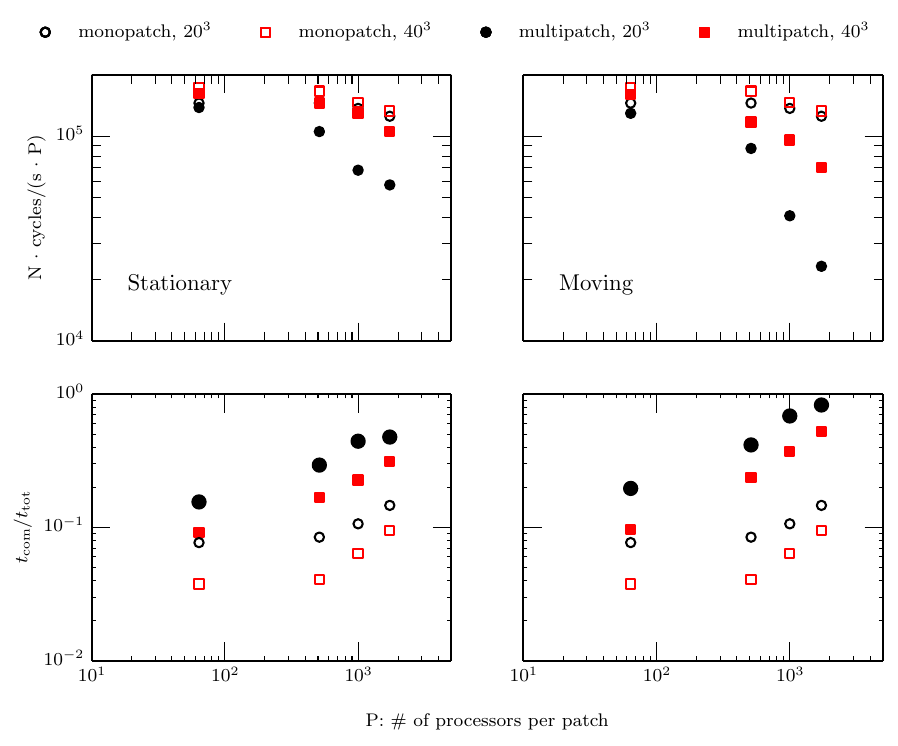}
\caption{Computational efficiency as a function of numbers of processors
  per patch and for different numbers of cells per processor.
  Monopatch method data are plotted with open symbols, multipatch with
  filled symbols.  Runs with $20^3$ cells per processor are shown with
  black circles, runs with $40^3$ cells per processor with red squares.
  Left (right) panels show multipatch simulations with a stationary (moving) patch.
  Top panels: Processing speed in zone-cycles per processor per second.
  Bottom panels: Fraction of total wall-clock time spent on communication.
\label{fig:comm}}
\end{figure*}

In our first set of benchmarks, we consider the case of a stationary
local patch and a single time-step for both patches.   The results
(obtained from STDLIB C ``time()")
are shown in the two left-hand panels of Figure~\ref{fig:comm}.
In this set, we assign the same number of processors to each patch;
that means both patches have the same number of cells per processor.
When the number of cells per processor is relatively
large, multipatch operations create a very modest overhead: the
ratio of cycle-update speed for multipatch to monopatch with $40^3$
cells per processor is $\simeq 0.75$.  On the other hand, with fewer
cells per processor ($20^3$), the ratio is closer to $\simeq 0.4$.

Another view of stationary patch computational efficiency may be
seen in the lower left panel of Figure~\ref{fig:comm}, showing
the ratio of time spent in communication relative to the total
computational time.   Here we define ``communication" as any
operations involving boundary condition exchange between processors.
Examples of communication specific to multipatch operation 
include determination of client-server relations, transmission
of ghost-cell coordinates, and interpolation of data to those coordinates.
A fourth category of communication, data transmission from server to client,
occurs in any sort of parallelized simulation. ``Total time" is defined as all time spent on
communication plus time spent on computing hydrodynamic updates;
it does not include time spent on ancillary activities such as
initialization or writing output.  In terms of this measure, we
find that for stationary patches the fraction of total time
spent in communication is $\simeq 2$ -- $3\times$ as great as for monopatch
runs with the same number of cells per processor.   This extra
time can be largely attributed to the interpolation step because
communication time in monopatch runs is due only to MPI data transmission,
whereas in multipatch operation it also includes interpolation and
client-router-server data exchange.

In the second set of comparisons (right panels of Fig.~\ref{fig:comm}),
we examine what happens when the local patch moves.   In this case,
the contrast in zone-update rate is larger and is also a stronger
function of the number of processors per patch.  As before, larger
numbers of cells per processor yield greater efficiency.   With
$40^3$ cells per processor, the update rate for the multipatch is
$\simeq 0.7\times$ the monopatch rate for 512 processors per patch,
but declines to $\simeq 0.5$ for 1728 processors per patch.  The
corresponding figures for $20^3$ cells per processor are 0.6 and $0.2\times$
the monopatch rate.  The fraction of time spent in communication is
consistently 5 -- $6\times$ larger in moving patches than for the
corresponding monopatch case, nearly independent of the number of processors per patch.

The overhead and scaling behavior of moving patches differs from that of
stationary patches because, in addition to data interpolation, it is also
necessary to determine client-server relationships at each time-step and
to transmit fresh ghost-cell coordinate lists.  These additional tasks both
increase the total overhead and make parallelization scaling poorer.  The
underlying reason is that as the number of processors per patch grows, more
processors must be queried to determine the correct set of client-server
connections.  The relative load this imposes is larger when the number
of cells per processor is smaller because, just like all other boundary
data considerations, it is sensitive to the processor domain's surface/volume
ratio.

We close this part of the discussion by making an important remark regarding interpretation of these
benchmarking results.  Although we expect the qualitative trends to be
robust, their quantitative character is sensitive both to the specific
problem and to the specific architecture of the computing system used.
Different problems (and different algorithms applied to the same problem)
can have different numbers of arithmetic operations per cell-update,
while different cluster architectures can give different rates of
inter-processor data transmission.  Because part of the multipatch
overhead depends on additional data transmission, the relative speeds
for these two sorts of processes can alter the multipatch/monopatch
comparison at a quantitative level.   As we discuss below (Sec.~\ref{sec:concl}),
technical improvements in implementation of the multipatch method can
also lead to quantitative changes in efficiency.

\begin{figure}[ht!]
\figurenum{15}
\plotone{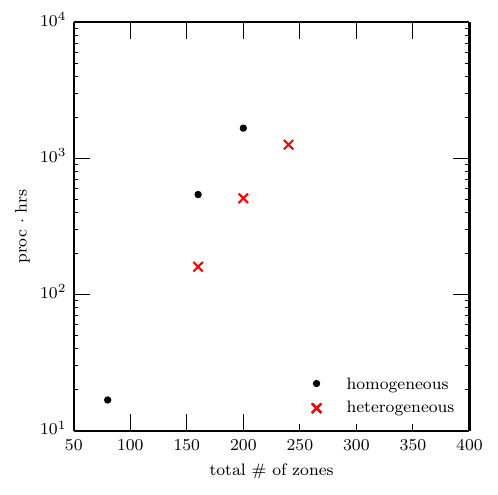}
\caption{ 
\label{fig:comm_SU}
Computational expense of a multipatch simulation with a homogeneous
(blue circles) and heterogeneous (orange X's) time-step for a given
total number of zones per patch $N$.  }
\end{figure}

Next, we compare the computational expense for a multipatch program
running with and without the heterogeneous time-step algorithm at
fixed total number of zones in each patch $N$.  The time-step ratio between the two
patches is $\Delta t_g / \Delta t_l \sim 8$.  For ideal load
balancing with a heterogeneous time-step, the global patch should therefore
have 8 times as many zones per processor as the local patch.  In
Figure \ref{fig:comm_SU}, we show how the number of processor-hours
required to reach a fixed physical time depends on the number of
zones per patch for both a homogeneous and a heterogeneous time-step.
As one might expect, both scale very closely to linearly with the
number of zones per patch, but the heterogeneous time-step requires
a number of processor-hours 2 -- $3\times$ smaller than homogeneous
time-step operation.

Note that the gains achieved by use of heterogeneous time-steps appear
in a way quite distinct from typical gains in efficiency.  They do not
lead to any increase in zone-cycles per processor per second; instead they
lead to a decrease in the total number of zone-cycles required to accomplish
the simulation.  In this way, their effect resembles the use of non-uniform
spatial grid cells, a device leading to economies in the total number of zones computed
by concentrating them where they are most needed, rather than a conventional
increase in computing efficiency.

\section{Conclusions}\label{sec:concl}

We have presented the essential methods underlying our implementation of
a new multipatch infrastructure, \Patch, designed to support multiscale, multiphysics, and
multireference frame fluid simulations.  This method offers a number of
advantages for the numerical study of complex fluid problems involving sub-regions
with contrasting properties.  Each patch can have its own coordinate system and
spatial grid, differing in geometry and resolution from all the other patches
(one of the many ways this can be useful is that if the  coordinate system preferable
for a part of the problem contains coordinate singularities, they can be covered with a new patch).
If different regions demand contrasting time-steps, the independence of the
processes evolving the patches permits them to have separately-determined time-steps,
potentially saving significant amounts of computing.  Although the method assumes
that a fluid exists throughout the problem volume, if different auxiliary
processes are important in different regions (e.g., chemical reaction networks
or self-gravity), their patches can treat those processes without burdening
the other regions.  Lastly, but possibly most importantly, substructures within
the problem may have differing preferred reference frames; these, too, can be
accommodated easily.

The patches are linked to one another solely through boundary condition exchange.
Contrasting grid systems are reconciled through interpolation; contrasting
grid geometries and reference-frames are reconciled through coordinate transformations
and ensuring that all transformed physical quantities are well-defined scalars,
vectors, or tensors.

Parallelization is essential to modern large-scale computing.  Arranging exchange
of boundary condition information between the correct processors can be a complex
problem in a multipatch system when the patches move relative to one another.  We
have constructed a solution to this problem---a client-router-server framework---that
updates these connections efficiently.  When the patches are stationary relative to
one another, the connections need to be identified only once, so the overhead due
to multipatch operations is fairly small, especially for larger numbers of cells
per processor.   When they move, the overhead is more significant and scales with the
number of processors per patch, producing a reduction in cell-update rate of about a
factor of $\sim 1.4$ for 512 processors per patch or a factor of $\sim 2$ for 1728
processors per patch when using $40^3$ cells per processor.  We note, however, that
these comparisons assume that monopatch and multipatch approaches use the same total
number of cells; because multipatch operation permits tuning the grid to match local
requirements, in practice multipatch simulations may use a much smaller total number
of cells than would be required for a monopatch simulation of the same problem---if
a monopatch simulation could deal with the problem at all.

Many extant fluid codes are automatically consistent with this infrastructure.
Its sole substantive stipulation is that the dependent variables involved in boundary
data exchange should be consistent
in all patches.  Although we were motivated to build this system by relativistic problems, and
our transformation methods are familiar because of their frequent application to
relativistic dynamics, in fact they really stem from more general considerations of
differential geometry; they therefore apply to any context in which scalars, vectors,
and tensors can be defined.

\Patch may be refined and extended, both in terms of its
computational efficiency and the span of physical problems on which it
can be used.  Communication between local patches (as opposed to only
local-global communication) can substantially extend the dynamic range
of lengthscales treated.  The amount of time spent on interpolation
and inter-patch data transmission can be reduced by minimizing the
number of arrays transferred or by eliminating unnecessary steps in
the coordinate transformations.  Moving from an MPMD environment to
one in which a single program employs task-based parallelization will
permit dynamical processor reassignment, amplifying the economies in
total zone-cycles necessary to compute that accrue from the use of
heterogeneous time-steps.  Task-based parallelization may also improve
interpolation efficiency because, for any single time-step, only a
fraction of the processors assigned to an individual patch are
involved in inter-patch data exchange.  Another improvement will be to
add interpolation options that, over a broader range of circumstances,
more nearly conserve quantities that should be conserved such as mass
and momentum.  Given suitable patch resolution, our current default
method does not create significant errors, but it would be valuable to
create new schemes, more nearly conservative, that would permit
greater freedom in resolution choices.  Similarly, some special
devices will be necessary to extend our multipatch method to MHD
problems in a way that preserves divergence-free magnetic field.  We
are currently developing algorithms to achieve this and hope to report
on them in the not-too-distant future.

\section*{Acknowledgments}
We thank Matt Duez for an insightful conversation at the onset of our
work, and thank Mark Avara for a careful reading of the paper and comments.
R.~M.~C., J.~H.~K. and H.~S. were partially supported by NASA
grant NNX14AB43G and NSF grant AST-1028111.  R.~M.~C. and
J.~H.~K. also received support from NSF grant AST-11516299.
R.~M.~C. is supported by a Nicholas C. Metropolis Postdoctoral
Fellowship under the auspices of the U.S. Dept. of Energy, and
supported by its contract W-7405-ENG-36 to Los Alamos National
Laboratory.  S.~C.~N. received support from NSF grants AST-1028087,
ACI-1515969, and AST-1515982.

This research project used computational resources at the Maryland Advanced
Research Computing Center (MARCC). We also thank the NSF for providing XSEDE resources
on the Stampede cluster through allocation TG-MCA95C003. Additional resources were provided through
Blue Waters sustained-petascale computing NSF projects ACI-0832606, ACI-1238993, OCI-1515969,and 
OCI-0725070. Blue Waters is a joint effort of the University of
Illinois at Urbana-Champaign and its National Center for
Supercomputing Applications.

\software{\HARM: \cite{Noble2009}}

\appendix

Implementation of the \Patch　system is accomplished through a set of functions enabling
an existing grid-based hydrodynamics code to run multipatch simulations.  The main tasks of
these additional functions may be organized according to three categories: setting up data infrastructure;
performing boundary condition exchange between patches; and a number of other utilities
specific to multipatch operation (e.g., managing heterogeneous timesteps).

In addition to grouping the multipatch routines in terms of function, it is also convenient
to group the changes necessary to convert a conventional hydrodynamics code in terms of user
control and responsibility.  Here, too, there are three categories:
``permanent" routines, user-supplied routines, and selected additional function calls.
The routines we designate as ``permanent" are those that remain fixed in character, independent
of the application.  These routines
\begin{itemize}
\item initialize \Patch's data infrastructure;
\item carry out inter-patch boundary data exchanges including interpolation and vector transformation;
\item restart the simulation when a patch has been added or removed (\S 2.4);
\item move patches;
\item coordinate time-steps, making allowance when necessary for heterogeneous time-steps (\S 2.6).
\end{itemize}

Several routines must be supplied by users because they are specific to the problem.  Their
calling sequences must have a specified form, but their contents are up to the user.  
They
\begin{itemize}
\item specify the geometric character of the patch and the
transformation linking its coordinate system to the background coordinates;
\item  create initial condition data (note that it is the user's responsibility
to ensure that initial data near patch boundaries are consistent with initial data in adjacent patches);
\item implement additional physics (when needed);
\item initialize configuration and motion of the patch.
\end{itemize}

Lastly, we made our best effort to minimize modifications to the underlying fluid
code by implementing the \Patch system as a wrapper.  For example, the hydrodynamics
part of the code is almost never touched. However, a few modifications must be made.

The great majority of these have to do with introducing MPMD features in MPI communication-related
routines.  In particular, it is necessary to redefine the term ``global".  In conventional
parallelized codes, ``global" denotes the entire problem volume and includes all processors
associated with the run.  However, in multipatch operation, it is necessary to distinguish
the entire volume of a patch from the entire volume of the problem.  Making this distinction
means that the global MPI communicator in the underlying fluid code must be redesignated ``local"
so that it refers only to the processors associated with an individual patch.  In addition,
there are often a number of variables and functions with both ``local" and ``global" versions;
all of these need to be renamed (in a way chosen by the user) to distinguish truly local
(in the domain of an individual processor) from patch-global (throughout a single patch's volume)
to problem-global (covering the entire problem volume).

There are also a number of places where the fluid code must call one of
the \Patch routines, sometimes one of the permanent routines, sometimes one of
the user-supplied routines.  Their purpose is to:
\begin{itemize}
\item initialize \Patch infrastructure;
\item move the patch to its next location (when necessary, this is done at the end of
each time-step);
\item construct patch boundary data (this happens at initialization, restart, and at the
end of each time-step); 
\end{itemize}

\noindent In addition, if optional features are used (e.g., add/remove patches or heterogeneous
time-steps), the user must likewise insert calls for them at appropriate places.

To close, we note that output from each patch can be handled by whatever means the
fluid code for that patch uses.  If, however, the user wishes to merge the datasets,
it is up to the user to write the software to accomplish it; the way such merges are
done is very problem-specific.

\bibliography{ms} \bibliographystyle{apj}

\end{document}